\documentclass[reqno,a4paper]{amsart}
\usepackage{pstricks,epsfig}
\usepackage{amsmath,amsfonts,latexsym,amssymb}
\usepackage{verbatim,amsthm,curves,graphics}
\newpsobject{showgrid}{psgrid}{subgriddiv=1,griddots=5,gridlabels=6pt}

\begin{document}
\title[Klein-Gordon Field with Boundary Oscillator]{The Massive Klein-Gordon
        Field Coupled to a Harmonic Oscillator at the Boundary}
\maketitle
\begin{center}
A. George\footnote{ag160@garfield.elte.hu}

\textit{Department of Mathematics, The University of York,}\\
\textit{Heslington, York, YO10 5DD, England.}
\end{center}
\begin{abstract}
We consider the massive Klein-Gordon field on the half line 
with and without a Robin boundary potential.
The field is coupled at the boundary to a harmonic oscillator.
We solve the system classically and observe the existence of classical
boundary bound states in some regions of the parameter space. The system
is then quantized, the quantum reflection matrix and reflection cross
section are calculated. Resonances and Ramsauer-Townsend effects are
observed in the cross section. The pole structure of the reflection
matrix is discussed. 
\end{abstract}

The study of quantum fields with boundaries has been the
subject of much
work in recent years. There have been two main methods used to study
such systems;
integrable boundary field theory (see, for example, \cite{GZ,Sk}) and
boundary perturbation theory (see \cite{BBT2}, and references there in).
Both approaches have focused on situations where the boundary
is a non-dynamic object that does not contain its own internal degrees of
freedom. 
Recently some studies have been conducted
on the coupling of
both classical and quantum
fields to boundaries, or impurities, containing additional degrees of
freedom, for instance, dynamic boundary conditions have been studied
for the integrable sine-Gordon \cite{BLZ}-\cite{BK}, supersymmetric
sine-Gordon \cite{N}, free
fermion fields \cite{FLFM} and several authors have studied the coupling of
the massless Klein-Gordon field to oscillators, e.g.
\cite{Cho}-\cite{FhM}.
These models can be used in the study of such physical systems as excited atoms in cavities and
quantum wires containing impurities.
Boundary degrees of freedom are also of interest in the study of brane-bulk interactions
in braneworld universes, see for example, \cite{BBC}. 

In this paper we consider another system consisting of a field with a dynamic
boundary.
This `toy model' consists of a massive
Klein-Gordon field in 1+1 dimensions restricted to the left half line by
a boundary. The field is linearly
coupled to a harmonic oscillator at the boundary thus introducing the
additional degrees of freedom.
Although this model is entirely linear, and may therefore appear trivial
on first inspection, we will see it possesses several interesting,
non-obvious features. The full analysis of these features, which we are
able to undertake due to its linear nature, will help in
the interpretation of similar effects in other non-integrable models and
in systems where
perturbative methods cannot be applied.

In section 1 we review previous results for the massive Klein-Gordon field with
the Robin boundary condition. 
In section 2 we consider the massive Klein-Gordon field on the halfline
coupled to an harmonic oscillator at the boundary with a Robin boundary potential. 
The system is solved in the classical regime 
and boundary bound states are observed in some regions of the parameter space.
We show that the classical solutions of the
system can be decomposed into independent modes of oscillation.
We observe that the existence of boundary bound states combined with the
requirement that the Hamiltonian be bounded below restricts the possible
values of the parameters of the theory.
We then
quantize the field directly from the classical solutions expressed as a
superposition of the independent
modes and find the quantum reflection
matrices of the systems from the two point function of the field. 
From the reflection matrix we calculate the reflection cross section, 
resonances and Ramsauer-Townsend effects are observed for some ranges 
of the parameter spaces.
We discuss the pole structure of the quantum reflection matrices.
At the end of section 2 we consider the special case when the Robin boundary potential is absent.
Section 3 contains a discussion on the main results of this paper, 
and possible directions for future work in the area.

\section{The Massive Klein-Gordon Field with a Robin Boundary}

In this section we
consider an example of an exactly solvable
boundary field theory, the massive Klein-Gordon
field on the left half line with a Robin boundary potential. 
This system has been studied previously, see, for example, \cite{Ful,BBT2}.

\subsection{Classical System}

The Hamiltonian for this system
is given by
\begin{equation}
\label{RH}
H=\int_{-\infty}^0\left( \frac{1}{2}\pi(x,t)^2
+\frac{1}{2}\left( \partial_x\phi(x,t)\right)^2
+ \frac{1}{2}m^2\phi(x,t)^2\right)\ dx
+ \frac{1}{2} \lambda \phi(0,t)^2\ ,
\end{equation}
where $\phi(x,t)$ and $\pi(x,t)$ are the field and its conjugate momentum
which have the Poisson bracket relations $\{\phi(x,t),\pi(y,t)\}=\delta(x-y)$
The boundary coupling parameter $\lambda$ is assumed to be real.

From \eqref{RH} we find Hamilton's equations for the system\footnote{
     Here and throughout this paper the notation $\partial_x\phi(0,t)$
     should be read as the derivative with respect to $x$ of $\phi(x,t)$
     evaluated at $x=0$, i.e. a shortened notation for $\partial_x\phi(x,t)|_{x=0}$.
},
\begin{align}
\label{RHam1}
\partial_t\phi(x,t)=\{\phi(x,t),H\}=& \pi(x,t)\ ,\\
\nonumber
\partial_t\pi(x,t)=\{\pi(x,t),H\}=& \partial_x^2\phi(x,t)-m^2\phi(x,t)\\
\label{RHam2}
& -\delta(x)\left( \partial_x\phi(0,t) + \lambda\phi(0,t) \right)\ .
\end{align}
From \eqref{RHam1} and \eqref{RHam2} we find the Robin boundary condition
by requiring that $\pi(x,t)$ be continuous at the origin,
\begin{equation}
\label{Rbc}
\partial_x\phi(0,t)=-\lambda\phi(0,t)\ .
\end{equation}
From Hamilton's equations \eqref{RHam1} and \eqref{RHam2} and the boundary
condition \eqref{Rbc} we recover the usual equation of motion for the massive
Klein-Gordon field,
\begin{equation}
\label{mKGeqom}
\partial_t^2\phi(x,t)=\partial_x^2\phi(x,t)-m^2\phi(x,t)\ .
\end{equation}

The equation of motion \eqref{mKGeqom} and boundary condition \eqref{Rbc}
are satisfied by the 
`bulk' solutions,
\begin{align}
\nonumber
\phi(x,t)=&\int_0^\infty \left( \cos(\rho x) 
-\frac{\lambda}{\rho}\sin(\rho x) \right)\\
\label{robinsoln}
&\qquad\times
 \left( a(\rho)\cos(\omega_\rho t)  
+\frac{b(\rho)}{\omega_\rho}\sin(\omega_\rho t) \right)\ d\rho\ ,
\end{align}
where $a(\rho)$ and $b(\rho)$ are real functions of $\rho$ and $\omega_\rho=\sqrt{m^2+\rho^2}$.

As well as the `bulk'
solutions there can exist square integrable boundary bound state
solutions. Such solutions can
found by allowing the momentum, $\rho$, of the
bulk solutions, given in this case by \eqref{robinsoln},
to take imaginary values, $\rho=i\varrho$. Because the bulk solutions satisfy the boundary
condition \eqref{Rbc} and the equation of motion
of the bulk field \eqref{mKGeqom} the imaginary
momentum solutions will also satisfy these equations. However, such
solutions will, generally, cause the field to diverge as $x \to -\infty$
and so must be discarded\footnote{
   The field is required to be bounded over its domain, thus
   solutions which diverge on the left half line must be discarded.}.
However, for special values of $\varrho$ the divergence of the field
arising from terms proportional to $\cosh (\varrho x)$ and that arising from
terms proportional to $\sinh (\varrho x)$ cancel out on the left half line. At such values of
$\varrho$ there exists a valid boundary bound state.
Such a valid boundary bound state exists for the Robin boundary at
$\varrho=-\lambda$ provided $\lambda$ is negative\footnote{
   If $\lambda$ is positive there is no solution which decays on the
   left half line instead there is a solution which exponentially grows
   as $x\to-\infty$ which must be rejected.
}. The boundary bound
state solution is then
\begin{equation}
\label{robinbbs}
\phi(x,t)=e^{\varrho x} \left(a_\varrho \cos(\omega_\varrho t) 
+ \frac{b_\varrho}{\omega_\varrho}\sin(\omega_\varrho t)\right)\ ,
\end{equation}
where $a_\varrho$ and $b_\varrho$ are real amplitudes and $\omega_\varrho=\sqrt{m^2-\varrho^2}$. 
Note that there is only a boundary bound state for negative values of
$\lambda$, i.e., when the contribution of the boundary term to the
Hamiltonian \eqref{RH} is negative. Such boundaries are called
attractive, those for which the contribution of the boundary term to the
Hamiltonian is positive are called repulsive boundaries. Repulsive
boundaries cannot support boundary bound states of this form. Also note
that when $\varrho^2>m^2$ the value of $\omega_\varrho$ becomes
imaginary and the boundary bound state solution diverges as
$t\to\pm\infty$. The most general solution for the Robin boundary is
formed by adding the bulk solutions to the boundary bound state
solution, if one exists.

We can now try substituting these solutions into the Hamiltonian
\eqref{RH}. Let us first assume that $\lambda$ is positive, i.e. the
boundary is repulsive and there exists no boundary bound state. Taking
each term of the Hamiltonian in turn at $t=0$, substituting the
expressions for the bulk solutions \eqref{robinsoln} 
we find that
\begin{equation}
\label{RH2rep}
H=\int_0^\infty\frac{1}{2} \tilde{b}(\rho)^2 + \frac{1}{2}\omega_\rho^2
\tilde{a}(\rho)^2 \ d\rho\ ,
\end{equation}
where
\begin{align}
\label{Ratilde}
\tilde{a}(\rho) :&= \sqrt{\frac{\pi}{2}} a(\rho) 
\left( 1+\frac{\lambda^2}{\rho^2}\right)^{1/2}\ ,\\
\label{Rbtilde}
\tilde{b}(\rho) :&= \sqrt{\frac{\pi}{2}} b(\rho) 
\left( 1+\frac{\lambda^2}{\rho^2}\right)^{1/2}\ .
\end{align}

When $\lambda$ is negative we must also include the boundary bound state
solution when we are substituting into the Hamiltonian \eqref{RH}.
This results in an additional term appearing in \eqref{RH2rep}
corresponding to the energy contribution of the additional mode. The
Hamiltonian for the attractive Robin boundary is
\begin{equation}
\label{RH2att}
H=\int_0^\infty\frac{1}{2} \tilde{b}(\rho)^2 + \frac{1}{2}\omega_\rho^2
\tilde{a}(\rho)^2 \ d\rho + \frac{1}{2} \tilde{b}_\varrho^2
+\frac{1}{2}\omega_\varrho^2\tilde{a}_\varrho^2\ ,
\end{equation}
where $\tilde{a}(\rho)$ and $\tilde{b}(\rho)$ are given by \eqref{Ratilde} and
\eqref{Rbtilde} and
\begin{align}
\nonumber
&{\tilde{a}_\varrho:=\frac{a_\varrho}{(2|\lambda|)^{1/2}}}\ ,\\
\nonumber
&{\tilde{b}_\varrho:=\frac{b_\varrho}{(2|\lambda|)^{1/2}}}\ .
\end{align}
Note that if $\lambda^2>m^2$ then $\omega_\varrho^2$ is negative. This
would mean that the Hamiltonian \eqref{RH2att} would not be bounded
below and thus be unphysical. It is therefore a requirement on the
parameters of the theory that $\lambda^2\leq m^2$ for the attractive Robin
boundary, as the repulsive boundary has no such bound state there does not
exist an equivalent condition, the Hamiltonian is always bounded below.

\subsection{Quantum System}

As both the Hamiltonian for the repulsive \eqref{RH2rep} and the attractive Robin
boundaries \eqref{RH2att} are now written as an infinite sum of
harmonic oscillators we can quantize the field in the canonical fashion. For
each mode we have a creation and annihilation operator,
\begin{align}
\nonumber
&{\alpha^\dagger(\rho):= \sqrt{\frac{\omega_\rho}{2}}\left(
\tilde{a}(\rho)-i\frac{\tilde{b}(\rho)}{\omega_\rho}\right)}\ ,
&{\alpha(\rho):=\sqrt{\frac{\omega_\rho}{2}}\left(
\tilde{a}(\rho)+i\frac{\tilde{b}(\rho)}{\omega_\rho}\right)}\ ,\\
\nonumber
&{\quad\alpha^\dagger_\varrho:= \sqrt{\frac{\omega_\varrho}{2}}\left(
\tilde{a}_\varrho-i\frac{\tilde{b}_\varrho}{\omega_\varrho}\right)}\ ,
&{\alpha_\varrho:=\sqrt{\frac{\omega_\varrho}{2}}\left(
\tilde{a}_\varrho+i\frac{\tilde{b}_\varrho}{\omega_\varrho}\right)}\ .\quad
\end{align}
The `bulk' creation operators, $\alpha^\dagger(\rho)$, act on the vacuum to produce bulk
particles of the field. These particles are located far to the left of
the boundary at $t=-\infty$ with momentum $p=\rho$ they propagate to
the boundary from which they reflect. At $t=\infty$ these particles are
once again located far to the left of the boundary and have momentum
$p=-\rho$. The creation operator associated with the boundary bound
state, $\alpha^\dagger_\varrho$, acts on the vacuum to create a particle of the field with energy
lower than that of a stationary bulk particle, this particle is
trapped close to the boundary at all times.

The one-particle quantum reflection matrix, or $R$-matrix, 
for the massive Klein-Gordon field with a linear boundary can be
found directly from the two point
function of the quantized field,
\begin{align}
\nonumber
\langle 0 | \phi (x_2,t_2) \phi (x_1,t_1)  | 0 \rangle
&= \int_{-\infty}^{\infty} \frac{1}{4 \pi \omega_p} e^{-i p
(x_2-x_1)} e^{-i \omega_p (t_2-t_1)} \ dp \\
\nonumber
& + \int_{-\infty}^{\infty} \frac{1}{4 \pi \omega_p} R (p)
e^{-i p (x_2+x_1)} e^{-i \omega_p (t_2-t_1)} \ d p \\
\label{Rdef}
& + \text{boundary bound state contributions.}
\end{align}
The first integral on the right hand side of \eqref{Rdef} is the
standard two point function of the Klein-Gordon field. This corresponds
to a particle propagating between the space time points $(x_1,t_1)$ and
$(x_2,t_2)$. The second integral corresponds to a particle propagating
between the same two points via the boundary where it is reflected and
picks up the momentum dependent amplitude $R(p)$, which is
the equivalent to the $R$-matrix found using any
other commonly used definitions, see \cite{BBT1}.
The third term on the right of \eqref{Rdef} arises from any boundary
bound states of the system, such contributions will fall off to zero as
$x_1\to-\infty$ and $x_2\to-\infty$ due to the localization of the boundary
bound state particles close to the boundary. 

Using these creation and annihilation operators, the definition of
the quantum reflection matrix \eqref{Rdef}
and the expressions for the general bulk solutions of the field
\eqref{robinsoln} it is simple to calculate the $R$-matrix for the
Robin boundary to be
\begin{equation}
\label{RRmatrix}
R(p)=\frac{p-i\lambda}{p+i\lambda}\ .
\end{equation}
Let us consider the analytic structure of \eqref{RRmatrix}. There is
always a single pole located at $p=-i\lambda$. When $\lambda$ is negative
this pole appears in the upper complex half plane as shown in figure 1A
where the black circle indicates the pole and the white circle indicates
a node.
The pole occurs at the same value of imaginary momentum for which we
have observed a classical boundary bound state. The interpretation of
this pole is clear, it corresponds to the existence of a quantum
boundary bound state which is equivalent to the classical boundary bound
state. However, when $\lambda$ is positive, and the pole appears in the
lower complex half plane as shown in figure 1B, there is no classical
boundary state to which the pole can correspond and 
the pole does not correspond to a quantum state of the system either.

\begin{figure}
\begin{center}
\begin{pspicture}(-2,-2)(2,2)
\psline(-2,0)(2,0)
\psline(0,-2)(0,1.75)
\rput(1.75,1.75){$p$}
\psline(1.5,1.5)(1.5,2)
\psline(1.5,1.5)(2,1.5)
\rput(0,2){A}
\pscircle[fillstyle=solid,fillcolor=white](0,-1){0.15}
\pscircle[fillstyle=solid,fillcolor=black](0,1){0.15}
\end{pspicture}
\qquad
\begin{pspicture}(-2,-2)(2,2)
\psline(-2,0)(2,0)
\psline(0,-2)(0,1.75)
\rput(1.75,1.75){$p$}
\psline(1.5,1.5)(1.5,2)
\psline(1.5,1.5)(2,1.5)
\rput(0,2){B}
\pscircle[fillstyle=solid,fillcolor=white](0,1){0.15}
\pscircle[fillstyle=solid,fillcolor=black](0,-1){0.15}
\end{pspicture}\\
\caption{The analytic structure of the quantum reflection matrix for the
Robin boundary for (A)
$\lambda<0$ 
and (B) $\lambda>0$.}
\end{center}
\end{figure}
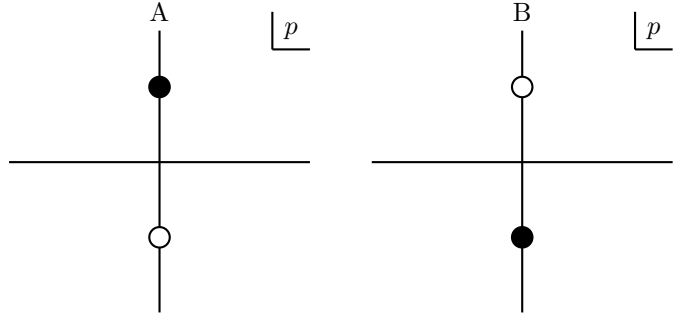  

\section{Coupling the Massive Klein-Gordon Field to a Boundary
Oscillator in the Presence of a Robin Potential}

In this section we consider the massive
Klein-Gordon field in 1+1 dimensions restricted to the left half line
with a Robin boundary potential. 
In addition the
field is linearly
coupled to a harmonic oscillator at the boundary thus introducing an
additional degrees of freedom.
This preserves the
linear, and thus exactly solvable, nature of the theory but introduces
new and interesting features into the reflection cross section.

\subsection{Classical System}
The Hamiltonian for this system is
\begin{align}
\nonumber
H=&\int_{-\infty}^{0}\left(\frac{1}{2}\pi(x,t)^2 
+ \frac{1}{2}\left(\partial_x\phi(x,t)\right)^2 
+ \frac{1}{2}m^2 \phi(x,t)^2\right)\ dx \\
\label{0hamiltonian}
&+ \frac{1}{2}\lambda \phi(0,t)^2 + \beta\phi(0,t)q(t)
+ \frac{1}{2}\mu^2q(t)^2 + \frac{1}{2}p(t)^2\ ,
\end{align}
where $\pi(x,t)$ is the conjugate momentum to the field $\phi(x,t)$, and
$p(t)$ is the conjugate momentum to the oscillator $q(t)$. These have
the usual equal time Poisson bracket relations,
$\{\phi(x,t),\pi(y,t)\}=\delta(x-y)$, $\{q(t),p(t)\}=1$, all other
brackets are zero. The parameters of the theory are the mass of the
field, $m$, the Robin boundary coupling parameter, $\lambda$, the boundary
oscillator coupling parameter, $\beta$, and the natural frequency of the
boundary oscillator, $\mu$. In natural units\footnote{
    Setting $c=\hbar=1$, gives relations between the dimensions,
    $[M]=[L^{-1}]=[T^{-1}]$.} 
these parameters have
dimensions $[m]=[\mu]=[\lambda]=[M]$, $[\beta]=[M^{3/2}]$ and for the
oscillator $[q(t)]=[M^{-1/2}]$ and $[p(t)]=[M^{1/2}]$. The
parameters $m$, $\mu$, $\lambda$ and $\beta$ are assumed to be real and
$m$ and $\mu$ are taken to be positive.

Hamilton's equations for the field and oscillator are
\begin{align}
\label{0hamphi}
\partial_t\phi(x,t)=&\ \{\phi(x,t),H\}=\pi(x,t)\ ,\\
\nonumber
\partial_t\pi(x,t)=&\
\{\pi(x,t),H\}=\partial_x^2\phi(x,t)-m^2\phi(x,t)\\
\label{0hampi}
&\ -\delta(x)\left(\partial_x\phi(0,t) + \lambda\phi(0,t) + \beta
q(t)\right)\ ,\\
\label{0hamq}
\partial_t q(t)=&\ \{q(t),H\}=p(t)\ ,\\
\label{0hamp}
\partial_t p(t)=&\ \{p(t),H\}=-\beta\phi(0,t)-\mu^2 q(t)\ .
\end{align}
From \eqref{0hampi} we find the boundary condition necessary for
$\pi(x,t)$ to be continuous to be
\begin{equation}
\label{0bc}
\partial_x\phi(0,t)=-\lambda\phi(0,t)-\beta q(t)\ .
\end{equation}
Combining the boundary condition \eqref{0bc} and Hamilton's equations
\eqref{0hamphi} and \eqref{0hampi} we recover the equation of motion for
the Klein-Gordon field
\begin{equation}
\label{0eqom}
\partial_t^2\phi(x,t)=\partial_x^2\phi(x,t)-m^2\phi(x,t)\ .
\end{equation}
Similarly we find the equation of motion for the boundary oscillator
from Hamilton's equations \eqref{0hamq} and \eqref{0hamp} to be
\begin{equation}
\label{0osceqom}
\partial_t^2q(t)=-\beta\phi(0,t)-\mu^2q(t)\ ,
\end{equation}

The equations of motion \eqref{0eqom} and \eqref{0osceqom} and boundary condition
\eqref{0bc} are satisfied by the bulk solutions,
\begin{align}
\nonumber
\phi(x,t)=&\ \int_0^\infty\left(\rho(\rho^2-\mu^2+m^2)\cos(\rho x)
-\left(\lambda(\rho^2-\mu^2+m^2)+\beta^2\right)\sin(\rho x)\right)\\
\label{0phisoln}
&\qquad \times\left(a(\rho)\cos(\omega_\rho t)+b(\rho)\sin(\omega_\rho
t)\right)\ ,
\end{align}
\begin{equation}
\label{0qsoln}
q(t)=\int_0^\infty \beta\rho\left(
a(\rho) \cos(\omega_\rho t) + b(\rho) \sin(\omega_\rho t) 
\right)\  d\rho\ ,
\end{equation}
Where $\omega_\rho=\sqrt{m^2+\rho^2}$ and $a(\rho)$ and $b(\rho)$ are
real functions of $\rho$.

These solutions have several interesting
features. For large $\lambda$ or $\beta$ the solutions for the field become equivalent,
after a rescaling of $a(\rho)$ and $b(\rho)$,
to solutions for the Dirichlet boundary condition and for small
$\lambda$ and $\beta$ the field solutions become the same as for a Neumann
boundary. 
These qualitative features of
the bulk solutions are shared by the solutions for the Robin boundary
condition discussed section 1, this behaviour is said to
`extrapolate between the Dirichlet and Neumann boundary conditions'.
The solutions for the dynamic boundary
\eqref{0phisoln} and \eqref{0qsoln} contain additional interesting features.
In the region of the parameter space where $\mu>m$
there exists a value of $\rho$,
$\rho=\sqrt{\mu^2-m^2}$ at which for any value of
$\beta$ and $\lambda$ the solution, for this one mode, becomes proportional to $\sin(\rho x)$, this can
be interpreted as a resonance effect, at this value of $\rho$ the field
is oscillating with frequency $\omega_\rho=\mu$, the natural frequency
of the boundary oscillator.
A second feature in the solutions not previously observed for
just the Robin boundary potential is
that when $\lambda\rho^2=\lambda(\mu^2-m^2)-\beta^2$ the single mode solution becomes
proportional to $\cos(\rho x)$. In the next section we will return to
these features when we consider the reflection cross section of this
boundary.

In addition to the spatially oscillating `bulk' solutions there exist square integrable `boundary bound state'
solutions found from the bulk solutions by allowing the parameter
$\rho$ to take imaginary values $\rho=i\varrho$ and requiring that the
divergence of the field on the left half line arising from the terms
proportional to $\cosh(\varrho x)$ and $\sinh(\varrho x)$ cancel.
These solutions have the form
\begin{align}
\label{0bbsphisoln}
\phi(x,t)=&\ e^{\varrho x}\left(a_\varrho\cos(\omega_\varrho t)
+b_\varrho\sin(\omega_\varrho t)\right)\ ,\\
\label{0bbsqsoln}
q(t)=&\ -\frac{\varrho+\lambda}{\beta}\left(a_\varrho\cos(\omega_\varrho t)
+b_\varrho\sin(\omega_\varrho t)\right)\ ,
\end{align}
where $\omega_\varrho=\sqrt{m^2-\varrho^2}$, $a_\varrho$ and $b_\varrho$
are real amplitudes and $\varrho$ is a positive real solution of
\begin{equation}
\label{0cubic}
\varrho^3 + \lambda\varrho^2 + (\mu^2-m^2)\varrho + \lambda(\mu^2-m^2) -
\beta^2 = 0\ . 
\end{equation}
We can split the parameter space of the theory into regions,
using the planes
\begin{itemize}
\item $\mu^2-m^2$=0
\item $\lambda=0$
\item $I=\lambda(\mu^2-m^2)-\beta^2=0$
\item $J_{\pm}=\frac{2}{3}\lambda(\mu^2-m^2)
       +\frac{2}{27}\lambda^3
       \pm\frac{2}{27}(\lambda^2-3(\mu^2-m^2))^{3/2}-\beta^2=0$
\end{itemize}
The value of $I$ is the value of the cubic $\varrho^3 + \lambda\varrho^2
+ (\mu^2-m^2)\varrho + \lambda(\mu^2-m^2) -\beta^2$ at $\varrho=0$,
$J_+$ is the value of the cubic at the local maximum of the function and
$J_-$ is the value of the cubic at the local minimum. If
$\lambda^2<3(\mu^2-m^2)$ there are no stationary points and $J_\pm$ take
complex values. When this occurs we use the real part of $J_\pm=0$ as the
boundary between regions. Figure 2 shows how the parameter space of the
theory is divided up by these planes.

\begin{figure}
\begin{center}

\vspace{1cm}
\begin{pspicture}(0,0)(12,8)
\psline(6,0)(6,7)
\rput(6,7.5){$\beta^2$}
\psline(1,0)(11,0)
\rput(11.5,0){$\lambda$}
\psline(6,0)(11,7)
\rput(11.5,7){$I=0$}
\pscurve(8,1.5)(9.5,5)(10,7)
\rput(10.25,7.25){$J_+=0$}
\pscurve(8,1.5)(9.5,4)(11,6.5)
\rput(11.75,6.5){$J_-=0$}
\pscurve[linestyle=dotted](8,1.5)(7,0.5)(6,0)
\rput(4,4){(a)}
\rput(8,5){(b)}
\rput(7.65,3.25){(d)}
\psline{->}(7.8,3.2)(8.25,2.75)
\rput(8.75,0.75){(f)}
\rput(9,6){(c)}
\psline{->}(9.15,6)(10,6)
\rput(10.5,4.75){(e)}
\psline{->}(10.35,4.75)(9.7,4.75)
\rput(0.5,7.5){A}
\rput(1.6,7.5){$\mu>m$}
\end{pspicture}\\

\vspace{1cm}
\begin{pspicture}(0,0)(12,8)
\psline(6,0)(6,7)
\rput(6,7.5){$\beta^2$}
\psline(1,0)(11,0)
\rput(11.5,0){$\lambda$}
\rput(0.5,7.5){B}
\rput(1.6,7.5){$\mu<m$}
\psline(6,0)(1,7)
\rput(0.5,6.75){$I=0$}
\pscurve(1.2,7)(6,1)(8,0)(10,1)(11,2.4)
\rput(11.35,2.6){$J_+=0$}
\rput(3.5,6){(g)}
\rput(5,2.75){(h)}
\psline{->}(4.8,2.55)(4.5,2.3)
\rput(3.5,1.25){(i)}
\rput(8,3){(j)}
\rput(7,1){(k)}
\psline{->}(6.8,0.8)(6.3,0.4)
\rput(10.5,0.5){(k)}
\end{pspicture}\\

\vspace{1cm}
\caption{The various regions of the parameter space of
the dynamic boundary with Robin potential theory for (A) $\mu>m$ and (B) $\mu<m$.}
\end{center}
\end{figure}

It is simple to find how many real solutions of \eqref{0cubic} exist in
each of these regions. For $\mu>m$ there is a single positive real
solution if $I<0$ which corresponds to regions (a), (b) and (c) in
figure 2A, if $J_+<0$, i.e. region (c), there are also two negative
real solutions to \eqref{0cubic}. If $\mu>m$ and $I>0$ then there are no
positive real solutions, but there is a single real negative solution in
regions (d) and (f) and three negative solutions in region (e). When
$\mu<m$ there is a single positive solution to \eqref{0cubic} if $I<0$,
i.e. in regions (g), (h), (j) and (k) of figure 2B. In regions (h) and
(k) there are also two negative real solutions. When $\mu<m$ and $I>0$,
i.e. in region (i) there are two positive real solutions and one
negative real solution. As only positive real solutions of \eqref{0cubic}
correspond to classical boundary bound state solutions this means there
exist one such solution in regions (a), (b), (c), (g), (h), (j) and (k).
There are no boundary bound states in regions (d), (e) and (f) and there
are two classical boundary bound state solutions in region (i).

For \eqref{0bbsphisoln} and \eqref{0bbsqsoln} to describe time oscillatory
states $\varrho$ must be less than $m$. Using \eqref{0cubic} it is
possible to express this requirement as an inequality between the
parameters of the model,
\begin{equation}
\label{0fcond}
\frac{\beta^2}{\mu^2}-\lambda\leq m\ .
\end{equation}
If this condition is satisfied the boundary bound states of the system
are time oscillatory, otherwise they are solutions in which the field
diverges as $t\to\pm\infty$.

Classical boundary bound state solutions for the Klein-Gordon field have
previously been observed for the Robin boundary, as discussed in section
1, where the equivalent condition to \eqref{fcond} for the bound state to
be oscillatory is $\lambda\leq m$. It is worth noting that the
bound state only exists for the attractive Robin boundary, i.e.
$\lambda<0$. If $\lambda>0$ then the Robin boundary is
repulsive and no bound state exists.

We will now show that the solutions \eqref{0phisoln}-\eqref{0bbsqsoln} can
be written as a superposition of orthogonal independent modes of
oscillation. Let us define
\begin{align}
\label{0atilde}
\tilde{a}(\rho):=&\ \sqrt{\frac{\pi}{2}}a(\rho)\left(
\rho^2\left(\rho^2-\mu^2+m^2\right)^2
+\left(\lambda(\rho^2-\mu^2+m^2)+\beta^2\right)^2\right)^{1/2} ,\\
\label{0btilde}
\tilde{b}(\rho):=&\ \sqrt{\frac{\pi}{2}}b(\rho)\omega_\rho\left(
\rho^2\left(\rho^2-\mu^2+m^2\right)^2
+\left(\lambda(\rho^2-\mu^2+m^2)+\beta^2\right)^2\right)^{1/2} ,
\end{align}
\begin{align}
\label{0varatildea}
\tilde{a}_\varrho:=&\ a_\varrho
\left(\frac{\beta^2+2\rho(\rho+\lambda)^2}{2\beta^2\rho}\right)^{1/2}\ ,\\
\label{0varatildeb}
\tilde{b}_\varrho:=&\ b_\varrho \omega_\varrho
\left(\frac{\beta^2+2\rho(\rho+\lambda)^2}{2\beta^2\rho}\right)^{1/2}\ ,
\end{align}
\begin{align}
\label{0psipx}
\psi(\rho,x):=&\ \frac{\rho(\rho^2-\mu^2+m^2)\cos(\rho x)
-\left(\lambda(\rho^2-\mu^2+m^2+\beta^2\right)\sin(\rho x)}
{\sqrt{\frac{\pi}{2}}\left(\rho^2(\rho^2-\mu^2+m^2)^2
+\left(\lambda(\rho^2-\mu^2+m^2)+\beta^2\right)^2\right)^{1/2}},\\
\label{0chip}
\chi(\rho):=&\ \frac{\beta\rho}{\sqrt{\frac{\pi}{2}}
\left(\left(\rho^2(\rho^2-\mu^2+m^2\right)^2
+\left(\lambda(\rho^2-\mu^2+m^2)+\beta^2\right)^2\right)^{1/2}}\ ,
\end{align}
\begin{align}
\label{0psivarpx}
\psi_\varrho(x):=&\ e^{\varrho x}
\left(\frac{2\beta^2\rho}{\beta^2
+2\rho(\rho+\lambda)^2}\right)^{1/2}\ ,\\
\label{0chivarp}
\chi_\varrho:=&\ -\frac{\rho+\lambda}{\beta}
\left(\frac{2\beta^2\rho}{\beta^2+2\rho(\rho+\lambda)^2}\right)^{1/2}\ .
\end{align}
We can now write the general solutions in terms of the definitions
\eqref{0atilde}-\eqref{0chivarp} as
\begin{align}
\nonumber
\phi(x,t)=\int_{0}^{\infty}&\ \left(
\tilde{a}(\rho)\psi(\rho,x)\cos(\omega_\rho t)
+\frac{\tilde{b}(\rho)}{\omega_\rho}\psi(\rho,x)\sin(\omega_\rho t)
\right)\ d\rho\\
\label{0genphisoln}
&+\sum_{\varrho}\left(\tilde{a}_\varrho \psi_\varrho(x)\cos(\omega_\varrho t)
+\frac{\tilde{b}_\varrho}{\omega_\varrho}
\psi_\varrho(x)\sin(\omega_\varrho t)\right)\ ,\\
\nonumber
q(t)=\int_0^{\infty}& \left(
\tilde{a}(\rho)\chi(\rho)\cos(\omega_\rho t)
+\frac{\tilde{b}(\rho)}{\omega_\rho}\chi(\rho)\sin(\omega_\rho t)
\right)\ d\rho\\
\label{0genqsoln}
&+\sum_\varrho\left(
\tilde{a}_\varrho \chi_\varrho \cos(\omega_\varrho t)
+\frac{\tilde{b}_\varrho}{\omega_\varrho}\chi_\varrho\sin(\omega_\varrho
t) \right)\ .
\end{align}
The sum in \eqref{0genphisoln} and \eqref{0genqsoln} runs over all
positive real roots of \eqref{0cubic}.
The functions \eqref{0psipx}-\eqref{0chivarp} form an orthonormal set over
the half line plus the oscillator,
\begin{gather}
\label{0orthog1}
\int_0^{\infty}\psi(\rho_1,x)\psi(\rho_2,x)\ dx
+ \chi(\rho_1)\chi(\rho_2)=\delta(\rho_1-\rho_2)\ ,\\
\label{0orthog2}
\int_0^\infty\psi_{\varrho}(x)\psi(\rho,x)\ dx
+\chi_\varrho\chi(\rho)=0\ , \\
\label{0orthog3}
\int_0^\infty\psi_{\varrho_1}(x)\psi_{\varrho_2}(x)\ dx
+ \chi_{\varrho_1}\chi_{\varrho_2}=\delta_{\varrho_1\varrho_2}\ .
\end{gather}
Relation \eqref{0orthog1} holds as $\rho$ is always positive,
\eqref{0orthog2} requires that $\varrho$ is a root of \eqref{0cubic} and
\eqref{0orthog3} uses the relations between roots of a cubic
equation\footnote{
   Let the roots of the cubic $z^3+c_2z^2+c_1z+c_0$ be $z_1$, $z_2$ and
   $z_3$. These roots satisfy the relations $z_1z_2z_3=-c_0$,
   $z_1z_2+z_1z_3+z_2z_3=c_1$ and $z_1+z_2+z_3=-c_2$. 
}.
By substituting the general solutions \eqref{0genphisoln} and
\eqref{0genqsoln} into the Hamiltonian \eqref{0hamiltonian} and applying
the othoganality relations \eqref{0orthog1}-\eqref{0orthog3} we can
rewrite the Hamiltonian as an infinite sum
of independent harmonic
oscillators each corresponding to one mode of oscillation of the field\footnote{
   I.e. the integral over all bulk modes plus the sum over the fininte number of boundstate modes.
},
\begin{equation}
\label{0hamiltonian2}
H=\frac{1}{2}\int_0^\infty \tilde{b}(\rho)^2
+\omega_\rho^2\tilde{a}(\rho)^2\ d\rho
+\frac{1}{2}\sum_\varrho \tilde{b}_\varrho^2
+\omega_\varrho^2\tilde{a}_\varrho^2\ .
\end{equation}
Note that the Hamiltonian \eqref{0hamiltonian2} is bounded below only if
$\omega_\varrho$ is real, i.e. if all the boundary bound state modes are
oscillatory, which occurs only when \eqref{0fcond} is satisfied.
Thus we must require that the condition
\eqref{0fcond} be satisfied to have a physical system.

\subsection{Quantum System}

By using the definitions \eqref{0atilde}-\eqref{0chivarp} we have described
the general classical solutions for the
field as the linear superposition of independent modes of
oscillation.
The Hamiltonian in this basis can be written as a sum over an infinite
number of harmonic oscillators each corresponding to one of the modes of
oscillation \eqref{0hamiltonian2}. 
It is now simple to quantize the system, by
analogy to the quantization of the unbounded field we can define
creation and annihilation operators,
\begin{align}
\nonumber
&{\alpha(\rho):=\sqrt{\frac{\omega_\rho}{2}}\left(
\tilde{a}(\rho)-i\frac{\tilde{b}(\rho)}{\omega_\rho}\right)\ ,}
&{\alpha_\varrho:=\sqrt{\frac{\omega_\varrho}{2}}\left(
\tilde{a}_\varrho-i\frac{\tilde{b}_\varrho}{\omega_\varrho}\right)\ ,}\\
\nonumber
&{\alpha^\dagger(\rho):=\sqrt{\frac{\omega_\rho}{2}}\left(
\tilde{a}(\rho)+i\frac{\tilde{b}(\rho)}{\omega_\rho}\right)\ ,}
&{\alpha^\dagger_\varrho:=\sqrt{\frac{\omega_\varrho}{2}}\left(
\tilde{a}_\varrho+i\frac{\tilde{b}_\varrho}{\omega_\varrho}\right)\ .}
\end{align}
Where there is one pair of creation annihilation operators for each
allowed boundary bound state.
The operators have the usual commutation relations for creation and
annihilation operators,
$[ \alpha (\rho_1) , \alpha^\dagger(\rho_2) ] =
\delta (\rho_1-\rho_2)$, $[ \alpha_\varrho , \alpha^\dagger_\varrho] =
1$, and $[ \alpha (\rho),\alpha^\dagger_\varrho ] = [ \alpha_\varrho ,
\alpha^\dagger (\rho) ] = 0$.

Using these operators we can construct a Fock space, let the vacuum, $|0\rangle$,
be the state which vanishes when acted on by any of the annihilation operators,
$\alpha(\rho)|0\rangle=0$, $\alpha_\varrho|0\rangle=0$. The one-particle
`bulk' states are created by the action of $\alpha^\dagger(\rho)$
on the vacuum, $\alpha^\dagger(\rho)|0\rangle=|\rho\rangle$, and one
particle boundary states are created by the action of
$\alpha^\dagger_\varrho$ on the vacuum,
$\alpha^\dagger_\varrho|0\rangle=|\varrho\rangle$.
The one-particle bulk states, $|\rho\rangle$, represent a particle of
mass $m$ that
at $t=-\infty$ is located far to the left of the boundary and is
travelling with momentum $p=\rho$ towards the boundary. At time
$t=\infty$ this particle is again located far to the left but is
travelling away from the boundary with momentum $p=-\rho$. At some
finite time the particle has approached and reflected from the boundary.
Boundary states, $|\varrho\rangle$, represent particles which are
located close to the boundary at all times.
Annihilation operators act on the one-particle states to return the
vacuum or zero,
$\alpha(\rho_1)|\rho_2\rangle=\delta(\rho_1-\rho_2)|0\rangle$,
$\alpha_{\varrho_1}|\varrho_2\rangle=\delta_{\varrho_1\varrho_2}|0\rangle$.
Multi-particle states
can be constructed from the action of several creation operators on the
vacuum. We can define the number operator for both the bulk particle
states, $N(\rho):=\alpha^\dagger(\rho)\alpha(\rho)$, and for the
boundary states, $N_\varrho:=\alpha^\dagger_\varrho\alpha_\varrho$, from
which we can construct the renormalized Hamiltonian,
\begin{equation}
\label{0hamiltonian3}
\hat{H}:=\int_0^\infty \omega_\rho N(\rho)\ d\rho
+\sum_\varrho \omega_\varrho N_\varrho\ .
\end{equation}
This is equivalent to the classical Hamiltonian \eqref{0hamiltonian2}
up to a constant infinite term.
From equation \eqref{0hamiltonian3} and the definition of $\omega_\rho$ we see
that the renormalized Hamiltonian is
hermitian if and only if $\omega_\varrho$ is real. This is the same
requirement as that for the classical boundary bound states to be oscillatory
in time. It has already been shown that this condition is equivalent to
equation \eqref{0fcond}. Thus provided that the condition
\eqref{0fcond} between the parameters of the model hold the
renormalized Hamiltonian \eqref{0hamiltonian3} will be a hermitian operator.
If \eqref{fcond} does not hold the theory has an unstable
vacuum and cannot be quantized.

From the general solution for the field \eqref{0genphisoln} and the
definition \eqref{Rdef} we find the reflection matrix to be
\begin{equation}
\label{0Rmatrix}
R(p)=\frac{p\left(p^2-\mu^2+m^2\right)-i\left(\lambda(p^2-\mu^2+m^2)+\beta^2\right)}
{p\left(p^2-\mu^2+m^2\right)+i\left(\lambda(p^2-\mu^2+m^2)+\beta^2\right)}
\ .
\end{equation}
Note that $R(p)$ is unitary $R(p)^\ast R(p)=1$ and that as the parameter
$p$ is the momentum of incoming particles 
and as the boundary defines the right most limit of the field
all incoming particles will have positive momentum, thus it is
sufficient for us to consider just these values of $p$.

The reflection matrix \eqref{0Rmatrix} has three poles corresponding to the three complex 
roots of \eqref{0cubic}. These poles either occure at positive, purely imaginary, values of
momentum ($p=i\varrho$) in which case they correspond to the bound state of the system, or
they occure at values of momentum in the lower half complex plane.
To help interpret these poles we
define the total reflection cross section by analogy with the
optical theorem for bulk scattering,
\begin{equation}
\label{0crosssection}
\sigma_{\text{tot}}(\rho) \propto \frac{1}{2} \Im \left( Q(\rho) \right) 
= \frac{\left(\lambda(\rho^2-\mu^2+m^2)+\beta^2\right)^2}
{\rho^2\left(\rho^2-\mu^2+m^2\right)^2+\left(\lambda(\rho^2-\mu^2+m^2)+\beta^2\right)^2}\ ,
\end{equation}
where $Q(\rho)=(-i)(R(\rho)-1)$ and $\Im(Q(\rho))$ indicates that we are taking the imaginary part of
$Q(\rho)$. 
The total reflection cross section $\sigma_{\text{tot}}(\rho)$,
measures the probability that, during the reflection process, the reflecting particle
in state $|\rho\rangle$ is scattered into some intermediate state,
$|\varsigma\rangle$, which then decays back into the original state\footnote{
   The total cross section $\sigma_{\text{tot}}$ here does not have a direct
   clear interpretation due to quantum mechanical interference between
   scattering and non-scattering terms, we pursue this interpretation
   through analogy with the bulk case.
   }.
Figure 3A shows the shape of $\sigma_{\text{tot}}$ for the dynamic
boundary in the region of the parameter space where $\mu>m$, for the special case when $\lambda=0$.
For comparison figure 3B shows the shape of the total reflection cross section of
the Robin boundary.
Both reflection cross sections exhibit a
background effect where, for low momentum particles, the cross section
increases arising from
the attraction or repulsion of the boundary to the scattering bulk
particles.
\begin{figure}
\begin{center}
\epsfig{file=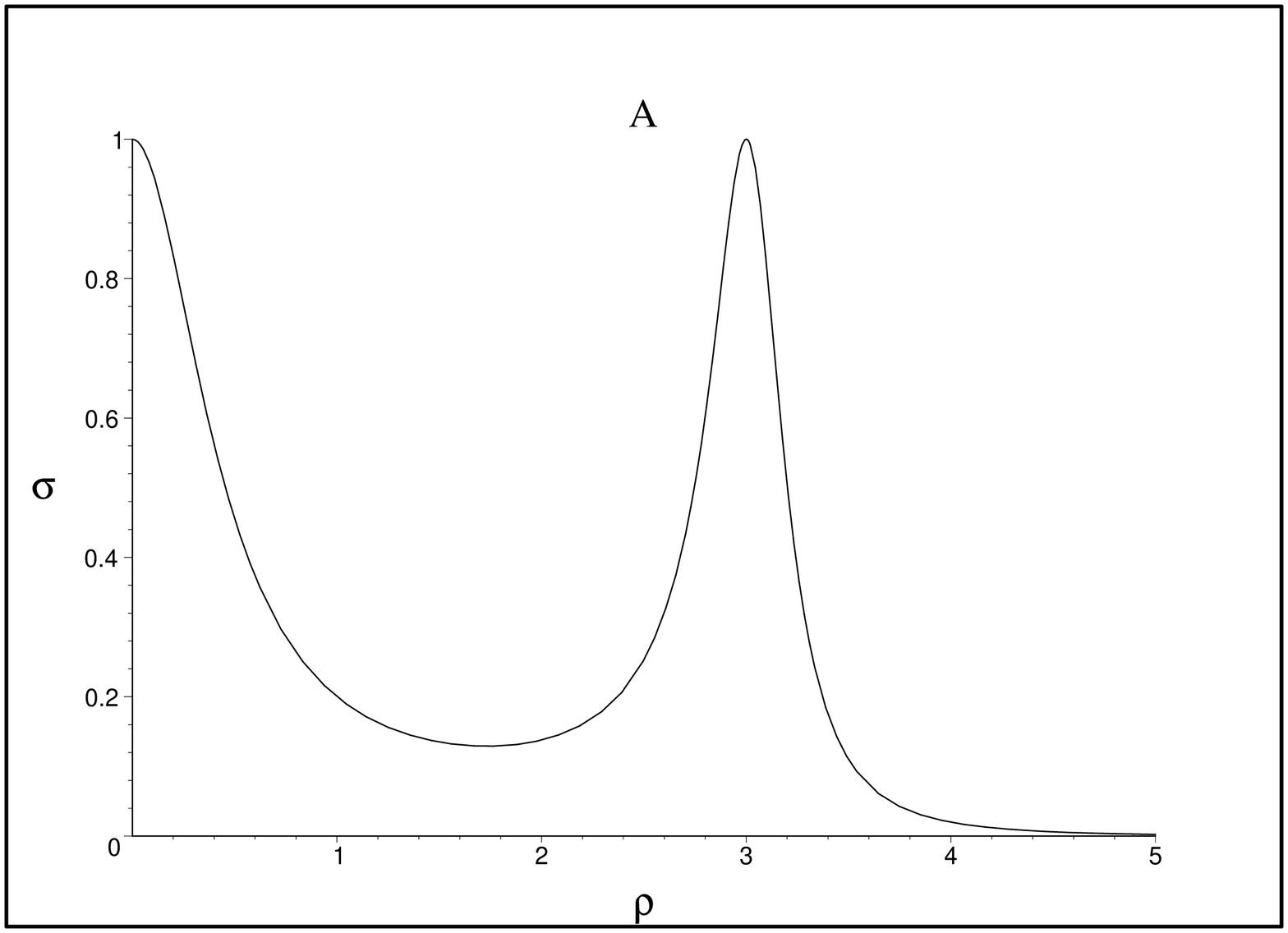, width=6cm, height=6cm, angle=270}
\epsfig{file=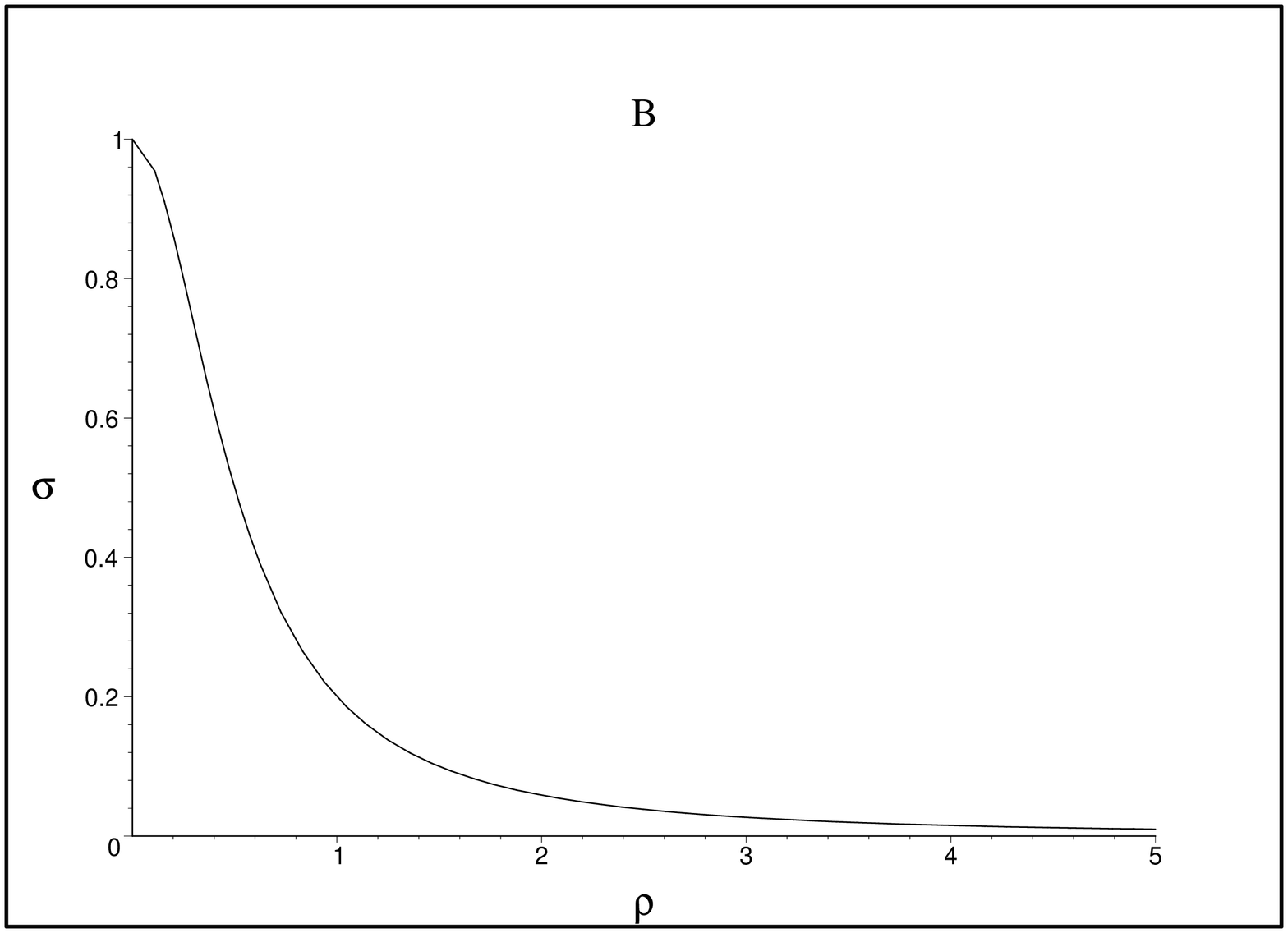, width=6cm, height=6cm, angle=270} \\
\caption{The total reflection cross sections of (A) the
dynamic boundary with $m=4$, $\mu=5$, $\beta=2$, $\lambda=0$ and (B) the repulsive
Robin boundary with $\lambda=\frac{1}{2}$.}
\end{center}    
\end{figure}
The
reflection cross section \eqref{0crosssection} has two other important
features, the first is a resonance peak around $\rho=\sqrt{\mu^2-m^2}$
where the cross section is large the second is a minimum where the cross
section falls to zero at $\lambda\rho^2=\lambda(\mu^2-m^2)-\beta^2$. These two features
correspond to when the classical solutions are proportional to
$\sin(\rho x)$ and $\cos(\rho x)$ respectively. 
Note that Figure 3A also exhibits a resonance peak but no minimum, we will discuss this
special case in a later section.
The resonance is only
observable in the reflection cross section when $\mu>m$ as $\rho$ is
positive and real for the scattering particles. Similarly the minimum
will only appear in the regions of the parameter space where $\mu>m$ and
$\lambda<0$, i.e.  
region (a) of figure 2A, or $\lambda>0$ and
$\beta^2<\lambda(\mu^2-m^2)$, regions (d), (e) and (f) of figure 2A. 
For
$\mu<m$ the minimum will appear if $\lambda<0$ and $\beta^2>\lambda(\mu^2-m^2)$,
regions (g) and (h)
of figure 2B. Figure 4 shows
the shape of the reflection cross section in the different regions of
the parameter space. Note the existence of the resonance peak in figures 4A,
4B and 4C and the zero cross section in 4A, after the resonance, 4B,
before the resonance, and in 4E.

\begin{figure}
\begin{center}
\epsfig{file=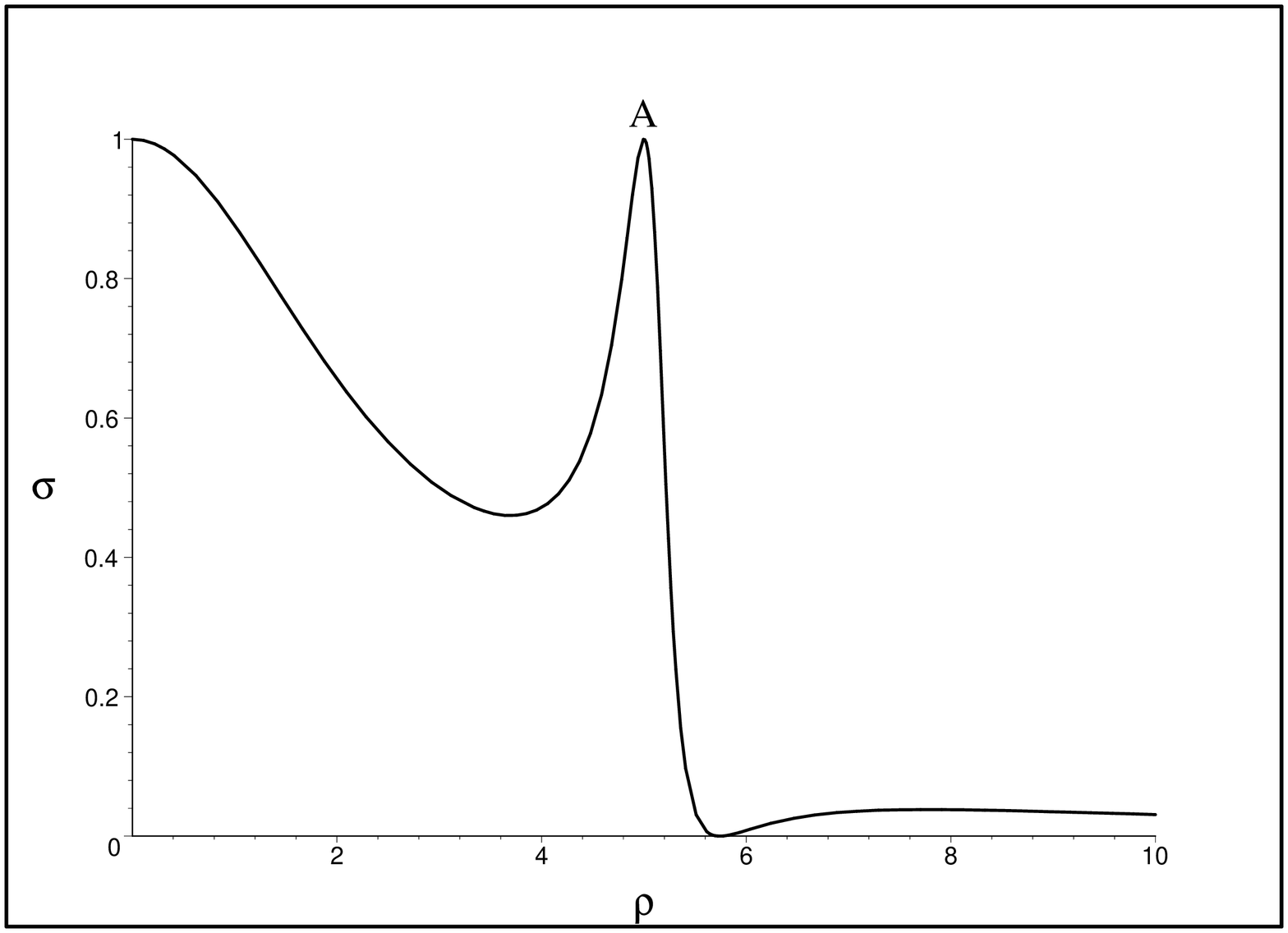,height=6cm,width=6cm,angle=270}
\epsfig{file=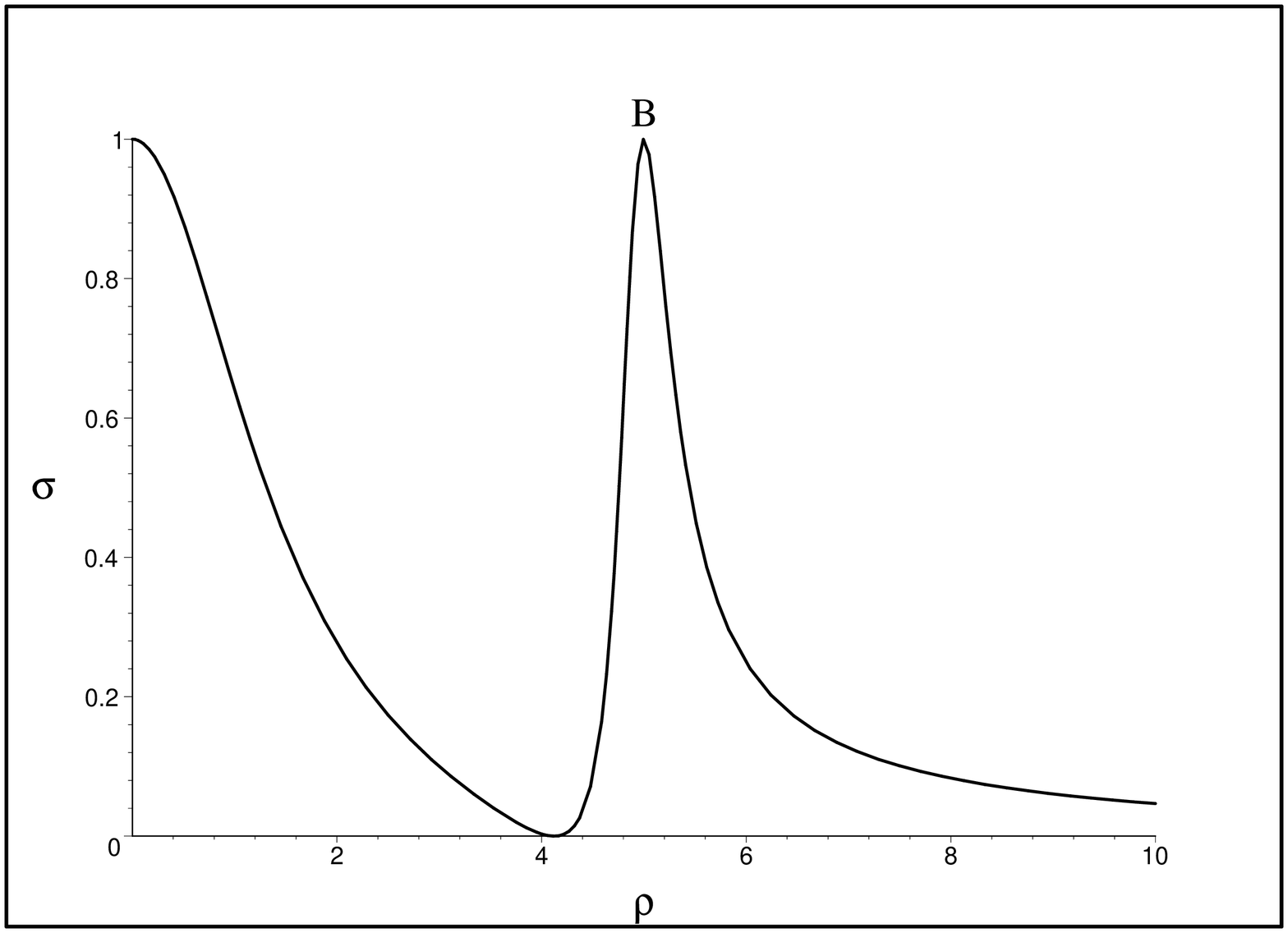,height=6cm,width=6cm,angle=270}\\
\epsfig{file=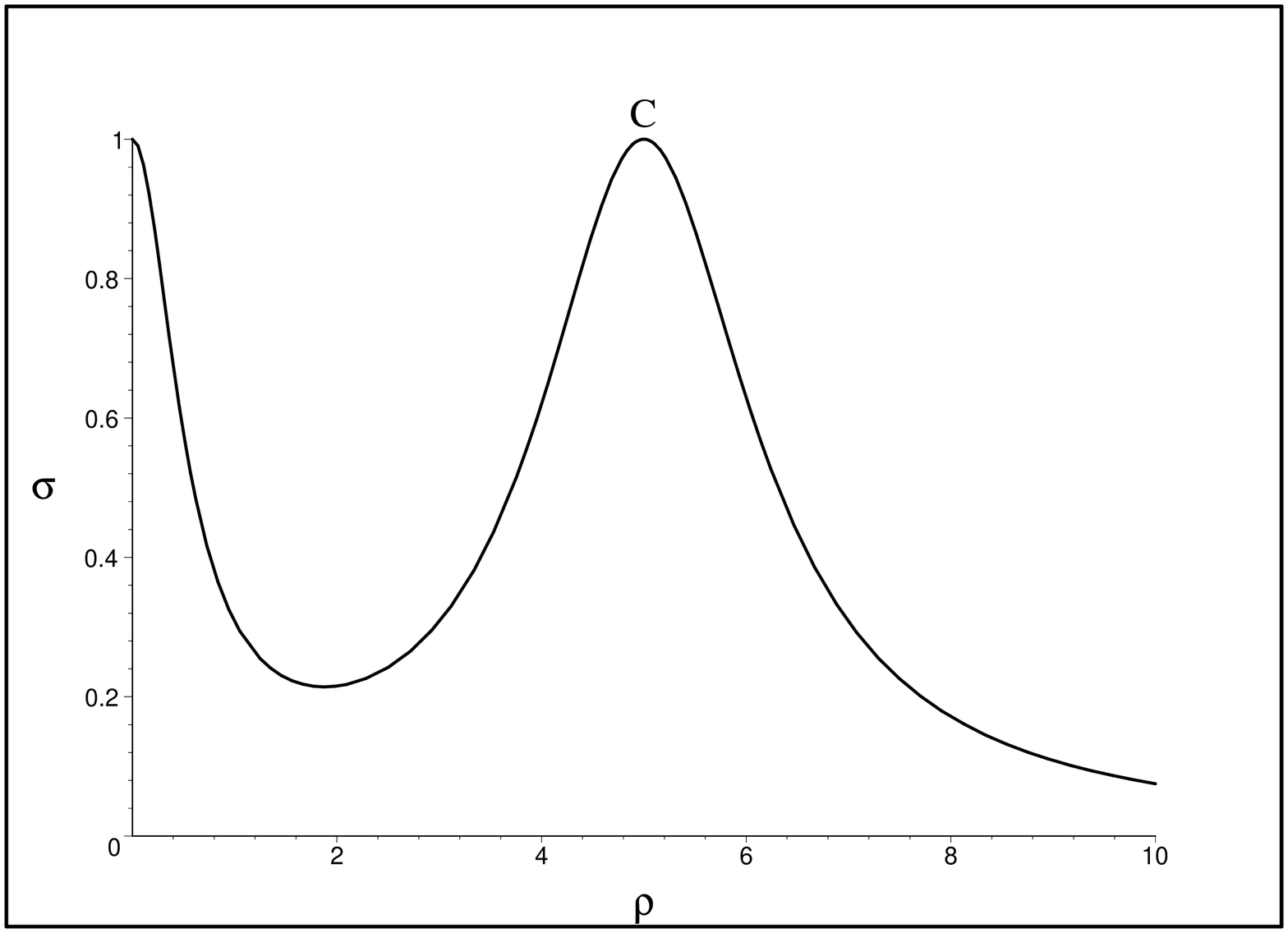,height=6cm,width=6cm,angle=270}
\epsfig{file=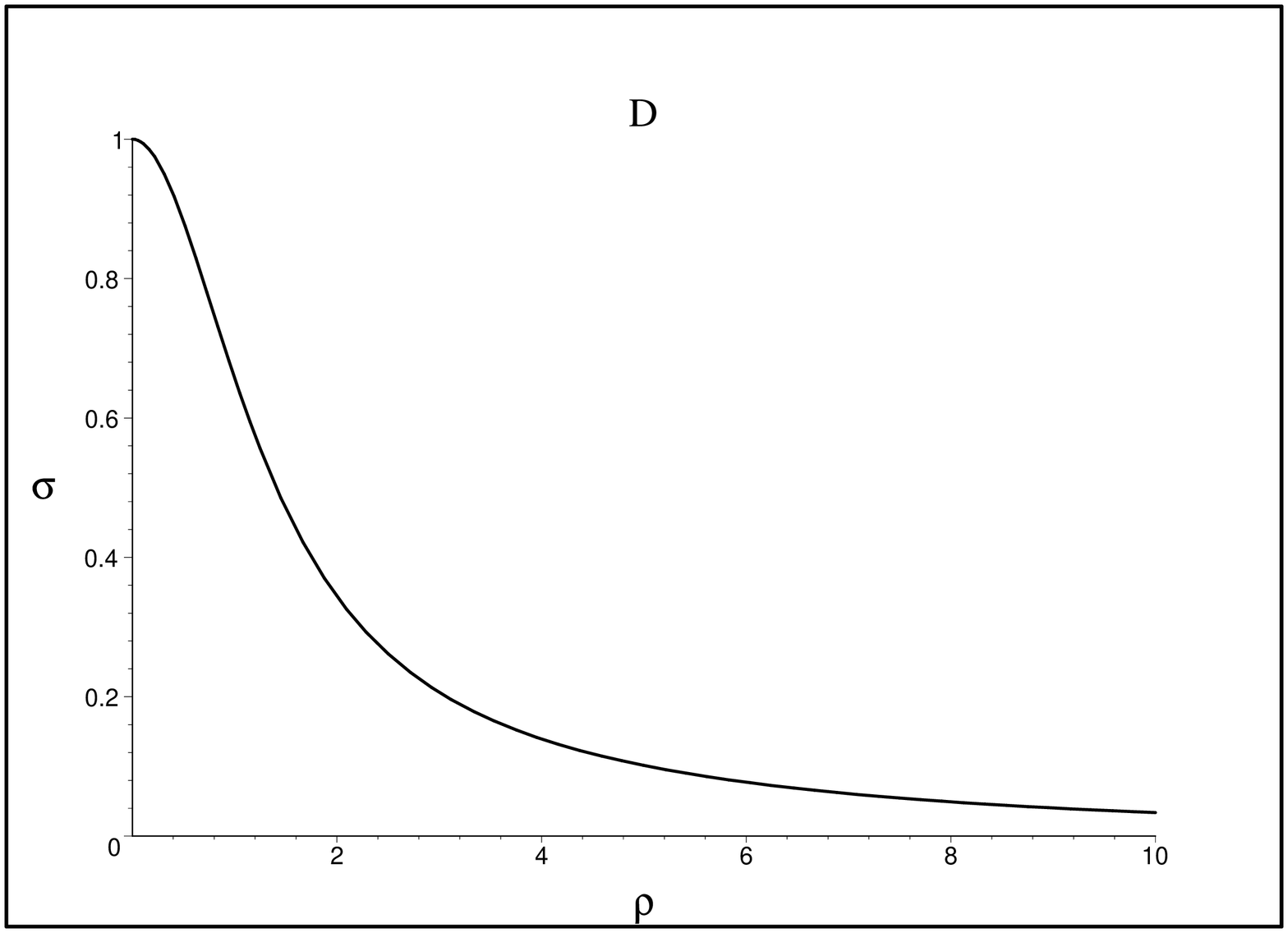,height=6cm,width=6cm,angle=270}\\
\epsfig{file=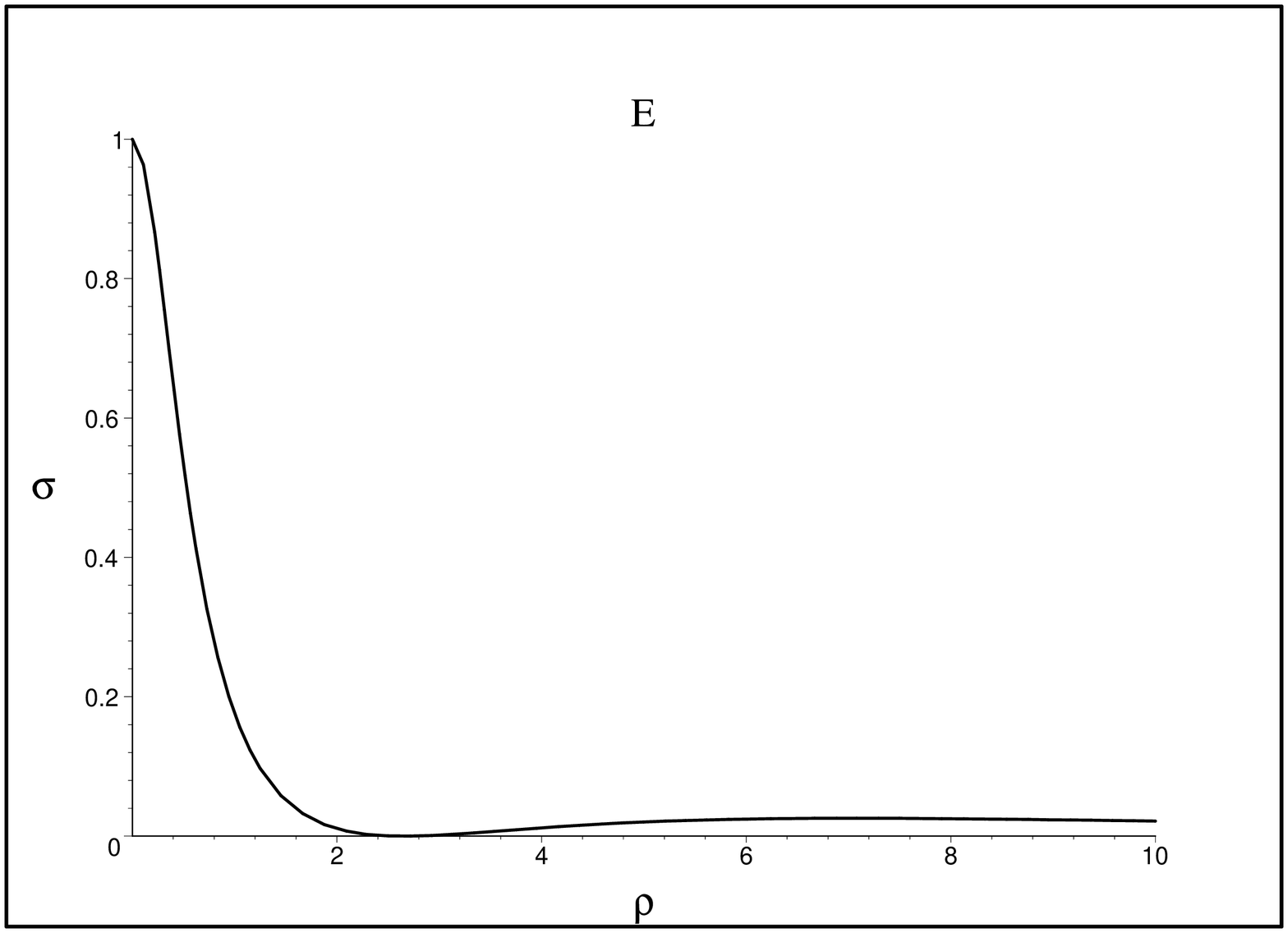,height=6cm,width=6cm,angle=270}
\epsfig{file=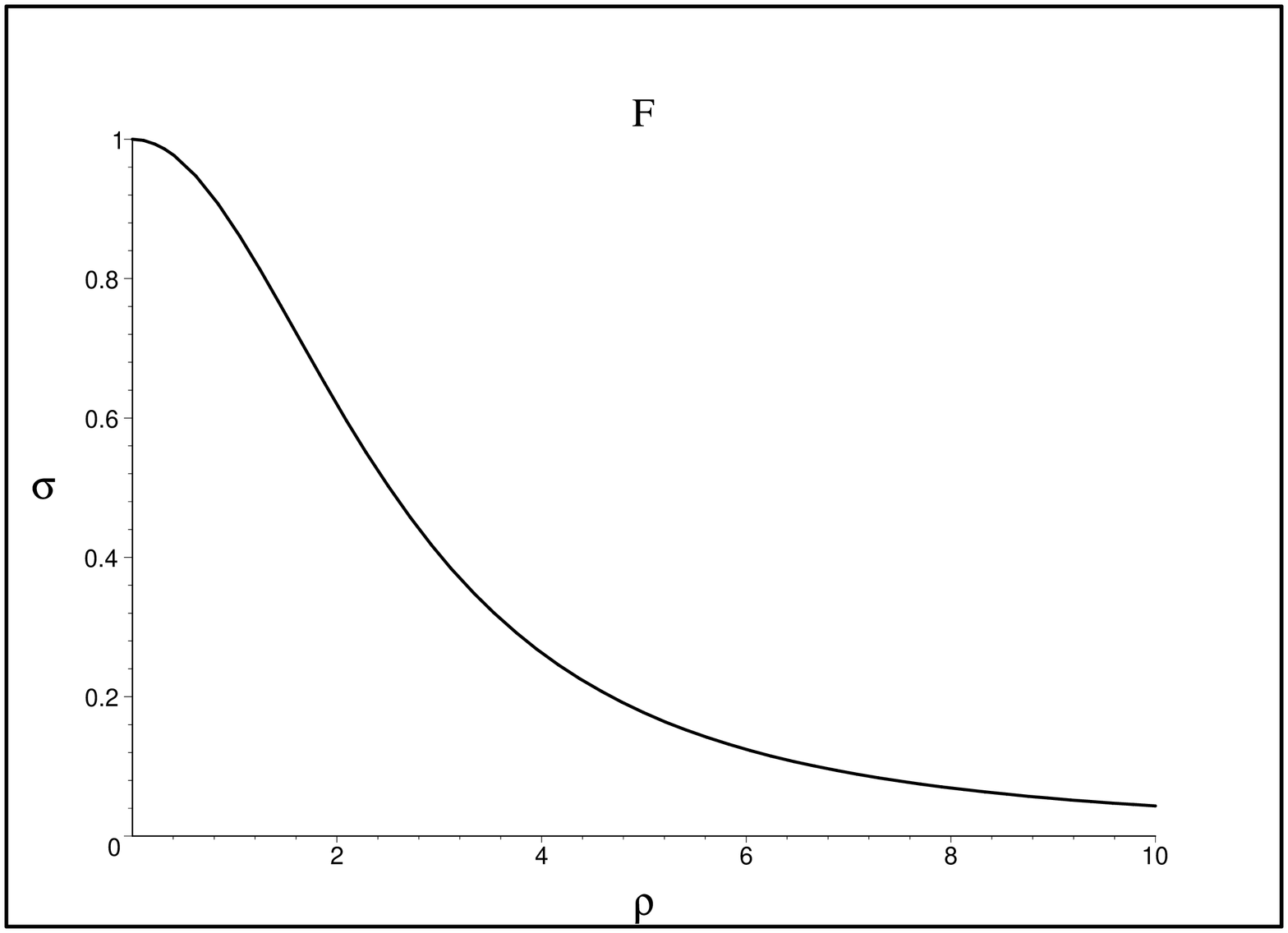,height=6cm,width=6cm,angle=270}\\
\caption{The total reflection cross section for the
boundary oscillator with Robin potential for (A) $\lambda=-2$, $\mu=13$,
$m=12$, $\beta=4$, (B) $\lambda=2$, $\mu=13$,
$m=12$, $\beta=4$, (C) $\lambda=2$, $\mu=13$,
$m=12$, $\beta=8$, (D) $\lambda=-2$, $\mu=12$,
$m=13$, $\beta=4$, (E) $\lambda=-2$, $\mu=12$,
$m=13$, $\beta=8$, (F) $\lambda=2$, $\mu=12$,
$m=13$, $\beta=4$.}
\end{center}
\end{figure}

The shape of the cross section close to a narrow resonance peak is well
approximated by the Breit-Wigner formula \cite{BW,W}
\begin{equation}
\label{BW}
\sigma_{\text{tot}}(\rho) \propto \left( (\omega_\rho - \omega_0 )^2 +
\frac{1}{4}\Gamma^2 \right)^{-1} \ ,
\end{equation}
where $\omega_\rho$ is the energy of the reflecting particle, $\omega_0$
is the mass of the resonance particle and $\Gamma$ is the decay
rate of the resonance.
By comparing \eqref{BW} and
\eqref{0crosssection} close to the resonance peak we find that the energy
of the resonance particle is equal to $\mu$, the natural frequency of the
boundary oscillator.
This suggests a new
interpretation for the resonance peak, 
if we view the boundary terms in the Hamiltonian, \eqref{0hamiltonian},
as describing a $0+1$ dimensional field rather than a mechanical
system then particles of this field will have mass $\mu$. Such particles
would be strongly
confined to the boundary as they are only able to propagate in the
domain of the boundary field. 
The resonance peak corresponds to a bulk
state, $|\rho\rangle$ with energy close to the mass, $\mu$, of the resonance
state, $|\varsigma\rangle$,
approaching the boundary and, through the linear coupling term, $\beta
\phi(0,t) q(t)$, creates a boundary particle in state $|\varsigma\rangle$ which
remains on the boundary
until it decays back into the bulk state $|\rho\rangle$.

From \eqref{BW} and \eqref{0crosssection} it is
also possible to calculate an approximate value for $\Gamma$ which will
be valid when the resonance is narrow,
\begin{equation}
\label{0gamma}
\Gamma \approx \frac{\beta^2}{\mu \sqrt{\mu^2-m^2}}\ .
\end{equation}
As $\mu$ approaches $m$ the shape of the resonance peak widens
corresponding to an increasing decay rate for the boundary particle.
The Breit-Wigner formula \eqref{BW} is only valid for narrow peaks so
\eqref{0gamma} ceases to be a good approximation.

Resonances associated with the boundary have been previously
observed in \cite{GZ,MMR,BPT,BBC}.

The zero reflection cross section arises from the combined effects
of the boundary oscillator and the Robin boundary
potential. A similar effect is observed in quantum mechanical scattering
from a square well potential, see, for example, \cite{Ra,Gr}. Such minima in
cross sections
are often associated with the Ramsauer-Townsend effect where scattering
of electrons from noble gas atoms show a minimum for certain energies of
scattering electrons \cite{Holt}.

We will now turn our attention to the pole structure of the quantum
reflection matrix \eqref{0Rmatrix}, taking each of the regions of the parameter
space in turn. In each case the reflection matrix has three poles
corresponding to the three complex solutions of the cubic \eqref{0cubic}.
First let us consider the region where $\mu>m$ and
$\lambda<0$, region (a) of figure 2A.  Figure 5 shows the analytic structure of the reflection
matrix in this region, black circles denote poles and white circles
denote nodes. In this case there in one pole on the positive imaginary
axis and two in the lower complex half plane.

\begin{figure}
\begin{center}
\begin{pspicture}(-3,-1.5)(3,1.5)
\psline(-3,0)(3,0)
\psline(0,-1.5)(0,1.5)
\rput(2.7,1.3){$p$}
\psline(2.5,1)(2.5,1.5)
\psline(2.5,1)(3,1)
\pscircle[fillstyle=solid,fillcolor=white](0,-1){0.15}
\pscircle[fillstyle=solid,fillcolor=black](0,1){0.15}
\pscircle[fillstyle=solid,fillcolor=white](-2,0.5){0.15}
\pscircle[fillstyle=solid,fillcolor=black](-2,-0.5){0.15}
\pscircle[fillstyle=solid,fillcolor=white](2,0.5){0.15}
\pscircle[fillstyle=solid,fillcolor=black](2,-0.5){0.15}
\end{pspicture}\\
\caption{The analytic structure of the reflection matrix of the dynamic
boundary with Robin potential
for $\mu>m$ and $\lambda<0$ (region (a)).}
\end{center}
\end{figure}
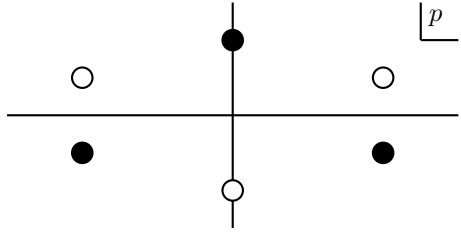

Earlier we observed a single classical
boundary bound state solution when $\mu>m$ provided
$\beta^2>\lambda(\mu^2-m^2)$. As $\beta^2$ and $(\mu^2-m^2)$ are positive
and $\lambda$ is negative there
exists a boundary bound state in this region which corresponds to the
pole on the positive imaginary axis in figure 5. Earlier in this section
we observed that the reflection cross section in this region of the
parameter space exhibits a resonance peak, as illustrated in figure
4A. The unstable resonance
state associated with this feature can be identified with the pole in
the lower right quadrant of figure 5. The third pole in the lower left
quadrant produces a resonance feature in the cross section at
$\rho=-\sqrt{\mu^2-m^2}$ however, as mentioned before, $\rho$ is a
positive parameter so this resonance is never observed. 

It is
illuminating to consider how the poles behave in the weak coupling
limit, i.e. as $\beta\to0$ and $\lambda\to0$ simultaneously. In this
case the pole corresponding to the bound state approaches $\rho=0$ along
the positive imaginary axis. This indicates that the state arises from
the stationary bulk particle state, the same behaviour as that of the
bound state of the attractive Robin
boundary. Such states can be interpreted as particles of the field with
energy less than their mass, $m$. These particles can only exist close to an
attractive boundary. Thus the boundary in this region of the parameter
space is attractive.
The poles corresponding to the resonance states of the system in the
weak coupling limit move onto the real $\rho$ axis at
$\rho=\sqrt{\mu^2-m^2}$. So when the coupling parameters are reduced
the resonance state stabilizes with energy equal to the mass of the
boundary field, $\mu$, as would be expected for a state which arises
from the particle state of the boundary field.

We will now consider the region of the parameter space where $\mu>m$,
$\lambda>0$ and $\beta^2<\lambda(\mu^2-m^2)$, regions (d), (e) and (f).
Figure 6 illustrates the
possible configurations of the poles of the reflection matrix in these
regions. In each case all three poles are located in the lower complex
half plane, and at least one pole is located on the imaginary axis.

\begin{figure}
\begin{center}
\begin{pspicture}(-2,-2)(2,2)
\psline(-2,0)(2,0)
\psline(0,-2)(0,1.75)
\rput(1.75,1.75){$p$}
\psline(1.5,1.5)(1.5,2)
\psline(1.5,1.5)(2,1.5)
\rput(0,2){A}
\pscircle[fillstyle=solid,fillcolor=black](0,-1.5){0.15}
\pscircle[fillstyle=solid,fillcolor=white](0,1.5){0.15}
\pscircle[fillstyle=solid,fillcolor=white](-0.75,0.5){0.15}
\pscircle[fillstyle=solid,fillcolor=black](-0.75,-0.5){0.15}
\pscircle[fillstyle=solid,fillcolor=white](0.75,0.5){0.15}
\pscircle[fillstyle=solid,fillcolor=black](0.75,-0.5){0.15}
\end{pspicture}
\
\begin{pspicture}(-2,-2)(2,2)
\psline(-2,0)(2,0)
\psline(0,-2)(0,1.75)
\rput(1.75,1.75){$p$}
\psline(1.5,1.5)(1.5,2)
\psline(1.5,1.5)(2,1.5)
\rput(0,2){B}
\pscircle[fillstyle=solid,fillcolor=black](0,-1.5){0.15}
\pscircle[fillstyle=solid,fillcolor=white](0,1.5){0.15}
\pscircle[fillstyle=solid,fillcolor=white](0,0.4){0.15}
\pscircle[fillstyle=solid,fillcolor=black](0,-0.4){0.15}
\pscircle[fillstyle=solid,fillcolor=white](0,0.9){0.15}
\pscircle[fillstyle=solid,fillcolor=black](0,-0.9){0.15}
\end{pspicture}
\
\begin{pspicture}(-2,-2)(2,2)
\psline(-2,0)(2,0)
\psline(0,-2)(0,1.75)
\rput(1.75,1.75){$p$}
\psline(1.5,1.5)(1.5,2)
\psline(1.5,1.5)(2,1.5)
\rput(0,2){C}
\pscircle[fillstyle=solid,fillcolor=black](0,-0.3){0.15}
\pscircle[fillstyle=solid,fillcolor=white](0,0.3){0.15}
\pscircle[fillstyle=solid,fillcolor=white](-0.8,1){0.15}
\pscircle[fillstyle=solid,fillcolor=black](-0.8,-1){0.15}
\pscircle[fillstyle=solid,fillcolor=white](0.8,1){0.15}
\pscircle[fillstyle=solid,fillcolor=black](0.8,-1){0.15}
\end{pspicture}\\
\caption{The analytic structure of the reflection matrix of the dynamic
boundary with Robin potential for $\mu>m$,
$\lambda>0$ and
(A) $\beta^2<J_-$ (region (f)) 
(B) $J_-<\beta^2<J_+<\lambda(\mu^2-m^2)$, or
$J_-<\beta^2<\lambda(\mu^2-m^2)<J_+$ (region (e)),
(C) $J_+<\beta^2<\lambda(\mu^2-m^2)$ (region(d)).
}
\end{center}
\end{figure}
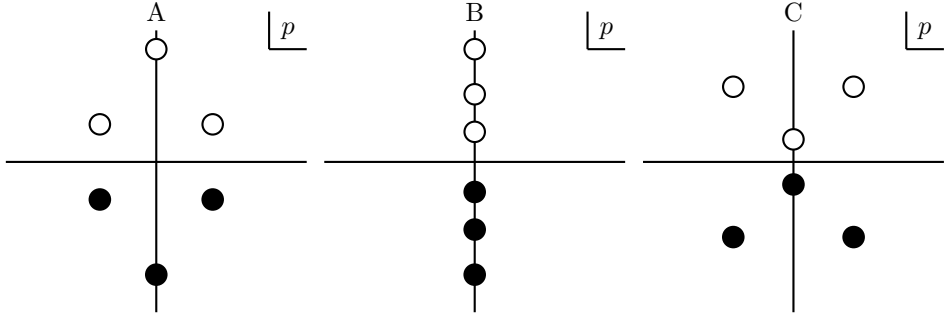

We know that there are no classical
boundary bound state solutions in this region of the parameter space,
poles located on the negative imaginary axis do not represent boundary
bound states like those located on the positive imaginary axis. We also
know that there is a resonance peak in this region. Let us consider the
behaviour of the poles in the weak coupling limit\footnote{
   where $\lambda$ and $\beta$ are reduced such that
   $\beta^2<\lambda(\mu^2-m^2)$ is always true.
}, one pole approaches $\rho=0$ along the negative imaginary axis, this
is the same weak coupling behaviour displayed by the pole for the repulsive Robin
boundary. Such poles do not carry a physical interpretation as a valid
state of the system. Boundaries whose reflection matrix contains poles
which have this weak coupling behaviour are repulsive to the
particles of the field.
The remaining poles of the reflection matrix approach
$\rho=\pm\sqrt{\mu^2-m^2}$ in the same way as described in the previous
case. As before we can interpret one of these poles as being the resonance
observed in figure 4B, even though it may be located at purely imaginary
values of $\rho$.
Neither of the other poles
of the reflection matrix have physical interpretations.

The third case we will consider is for $\mu>m$, $\lambda>0$ and
$\beta^2>\lambda(\mu^2-m^2)$, regions (b) and (c) of figure 2A. Figure
7 shows the possible configurations
of the poles of $R(\rho)$ in this region. Again we have a resonance as
seen in figure 4C and we also have a classical bound state solution,
which corresponds to the pole on the positive imaginary axis as usual.
If we take the low coupling limit, ensuring $\beta^2>\lambda(\mu^2-m^2)$
remains valid, we find that this pole approaches
$\rho=0$ along the positive imaginary axis indicating that the boundary
in this region is attractive.
We can consider how the state behaves if
we were to reduce $\beta^2$ and hold $\lambda$ constant (or increase
$\lambda$ and hold $\beta$ constant). In this case
the bound state pole passes through $\rho=0$ when
$\beta^2=\lambda(\mu^2-m^2)$. If $\beta^2$ is decreased further we will
pass into a region of the parameter space where the boundary is
repulsive. We know that this region does not have a bound state, thus we
conclude that the state corresponding to the bound state pole evaporates into a stationary
bulk particle state as we make the transition between the two regions.
The remaining poles behave in the low coupling limit in the same manner
as described in the previous two cases and one is associated with the
resonance state.

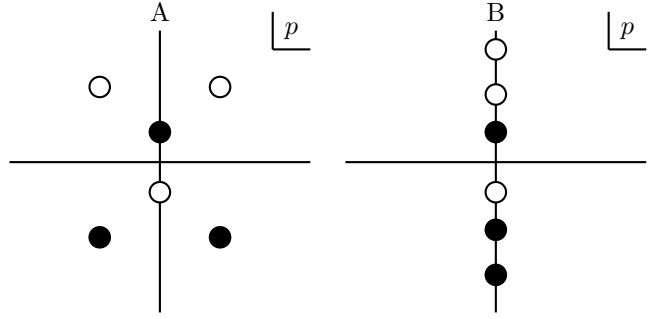
\begin{figure}
\begin{center}
\begin{pspicture}(-2,-2)(2,2)
\psline(-2,0)(2,0)
\psline(0,-2)(0,1.75)
\rput(1.75,1.75){$p$}
\psline(1.5,1.5)(1.5,2)
\psline(1.5,1.5)(2,1.5)
\rput(0,2){A}
\pscircle[fillstyle=solid,fillcolor=white](0,-0.4){0.15}
\pscircle[fillstyle=solid,fillcolor=black](0,0.4){0.15}
\pscircle[fillstyle=solid,fillcolor=white](-0.8,1){0.15}
\pscircle[fillstyle=solid,fillcolor=black](-0.8,-1){0.15}
\pscircle[fillstyle=solid,fillcolor=white](0.8,1){0.15}
\pscircle[fillstyle=solid,fillcolor=black](0.8,-1){0.15}
\end{pspicture}
\quad
\begin{pspicture}(-2,-2)(2,2)
\psline(-2,0)(2,0)
\psline(0,-2)(0,1.75)
\rput(1.75,1.75){$p$}
\psline(1.5,1.5)(1.5,2)
\psline(1.5,1.5)(2,1.5)
\rput(0,2){B}
\pscircle[fillstyle=solid,fillcolor=black](0,-1.5){0.15}
\pscircle[fillstyle=solid,fillcolor=white](0,1.5){0.15}
\pscircle[fillstyle=solid,fillcolor=black](0,0.4){0.15}
\pscircle[fillstyle=solid,fillcolor=white](0,-0.4){0.15}
\pscircle[fillstyle=solid,fillcolor=white](0,0.9){0.15}
\pscircle[fillstyle=solid,fillcolor=black](0,-0.9){0.15}
\end{pspicture}\\
\caption{The analytic structure of the reflection
matrix of the dynamic boundary with Robin potential for $\mu>m$, $\lambda>0$ and 
(A)$\beta^2>\lambda(\mu^2-m^2)>J_+$, 
or $\beta^2>J_+>\lambda(\mu^2-m^2)$ (region (b)),
(B)$\lambda(\mu^2-m^2)<\beta^2<J_+$ (region (c)).
}
\end{center}
\end{figure}

The fourth region we will consider is $\mu<m$, $\lambda<0$ and
$\beta^2<\lambda(\mu^2-m^2)$, region (i) of figure 2B, we know that in this region there are two
classical boundary bound state solutions, but there is no resonance
state. Figure 8 shows the arrangement of poles of $R(\rho)$ for this
region, as expected there are two poles on the positive imaginary axis
corresponding to the two bound states.

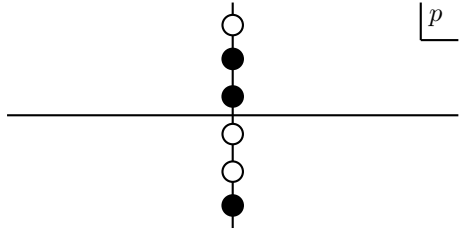
\begin{figure}
\begin{center}
\begin{pspicture}(-3,-1.5)(3,1.5)
\psline(-3,0)(3,0)
\psline(0,-1.5)(0,1.5)
\rput(2.7,1.3){$p$}
\psline(2.5,1)(2.5,1.5)
\psline(2.5,1)(3,1)
\pscircle[fillstyle=solid,fillcolor=white](0,1.2){0.15}
\pscircle[fillstyle=solid,fillcolor=black](0,-1.2){0.15}
\pscircle[fillstyle=solid,fillcolor=white](0,-0.75){0.15}
\pscircle[fillstyle=solid,fillcolor=black](0,0.75){0.15}
\pscircle[fillstyle=solid,fillcolor=white](0,-0.25){0.15}
\pscircle[fillstyle=solid,fillcolor=black](0,0.25){0.15}
\end{pspicture}\\
\caption{The analytic structure of the reflection matrix of the dynamic
boundary with Robin potential for $\mu<m$,
$\lambda<0$ and $\beta^2<\lambda(\mu^2-m^2)$ (region (i)).}
\end{center}
\end{figure}

In the weak coupling limit one of the bound state poles approaches
$\rho=0$ along the positive imaginary axis, indicating that this state
is a boundary bound state in the sense of an attractive Robin boundary bound
state, it arises from the stationary bulk particle state as
the coupling is turned on. The presence of this state also indicates
that the boundary is attractive in this region. 
The other poles behave differently in the weak coupling limit, they move
towards the purely imaginary values, $\rho=\pm\sqrt{\mu^2-m^2}$. The energy of these
states becomes the same as the mass of the boundary field, $\mu$.
Clearly these states arise from the particle state of the boundary field
in the same manner as the resonances do when $\mu>m$. 
The pole on the positive imaginary axis is interpreted as the stable boundary state
arising from the boundary field, the other pole lacks a physical
interpretation.

The final parameter space regions we need to consider is when $\mu<m$, and
$\beta^2>\lambda(\mu^2-m^2)$, which includes everywhere that
$\lambda>0$, the regions (g), (h), (j) and (k) of figure 2B. We know that in these regions there is a single classical
boundary bound state and no resonance state. Figure 9 shows the possible analytic
structure of the reflection matrix in these regions. In the weak
coupling limit the pole on the positive imaginary axis moves to
$\rho=\sqrt{\mu^2-m^2}$, and carries the interpretation of being the
state arising from the boundary particle state. A second pole approaches
$\rho=0$ along the negative imaginary axis, indicating that the boundary
is repulsive. 

\begin{figure}
\begin{center}
\begin{pspicture}(-2,-2)(2,2)
\psline(-2,0)(2,0)
\psline(0,-2)(0,1.75)
\rput(1.75,1.75){$p$}
\psline(1.5,1.5)(1.5,2)
\psline(1.5,1.5)(2,1.5)
\rput(0,2){A}
\pscircle[fillstyle=solid,fillcolor=white](0,-1.4){0.15}
\pscircle[fillstyle=solid,fillcolor=black](0,1.4){0.15}
\pscircle[fillstyle=solid,fillcolor=white](-0.8,0.5){0.15}
\pscircle[fillstyle=solid,fillcolor=black](-0.8,-0.5){0.15}
\pscircle[fillstyle=solid,fillcolor=white](0.8,0.5){0.15}
\pscircle[fillstyle=solid,fillcolor=black](0.8,-0.5){0.15}
\end{pspicture}
\qquad
\begin{pspicture}(-2,-2)(2,2)
\psline(-2,0)(2,0)
\psline(0,-2)(0,1.75)
\rput(1.75,1.75){$p$}
\psline(1.5,1.5)(1.5,2)
\psline(1.5,1.5)(2,1.5)
\rput(0,2){B}
\pscircle[fillstyle=solid,fillcolor=black](0,1.5){0.15}
\pscircle[fillstyle=solid,fillcolor=white](0,-1.5){0.15}
\pscircle[fillstyle=solid,fillcolor=black](0,-0.4){0.15}
\pscircle[fillstyle=solid,fillcolor=white](0,0.4){0.15}
\pscircle[fillstyle=solid,fillcolor=white](0,0.9){0.15}
\pscircle[fillstyle=solid,fillcolor=black](0,-0.9){0.15}
\end{pspicture}\\
\caption{The analytic structure of the reflection matrix of the dynamic
boundary with Robin potential for $\mu<m$
and (A) $\beta^2>J_+$, (regions (g) and (j)), and (B)
$\lambda(\mu^2-m^2)<\beta^2<J_+$ (regions (h) and (k)).}
\end{center}
\end{figure}
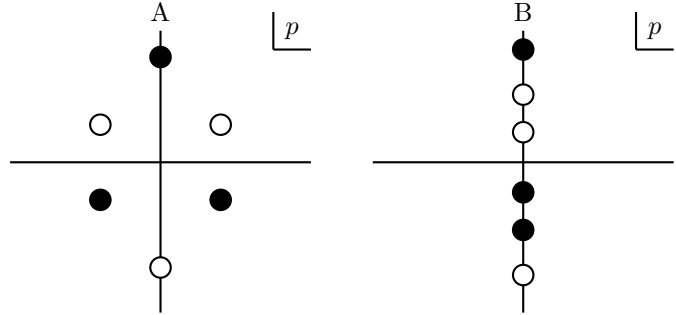

\subsection{The $\lambda=0$ special case}

In this section we will discuss a special case of the system considered previously where $\lambda=0$
i.e. where the Robin boundary potential is removed leaving the field just coupled to the oscillator
at the boundary.

Figure 10 shows the possible real solutions of \eqref{0cubic} for
$\lambda=0$ and for different values of
$m$, $\mu$ and $\beta$, clearly there is one and only one real positive
solution for any allowed choice of the parameters.

\begin{figure}
\begin{center}
\epsfig{file=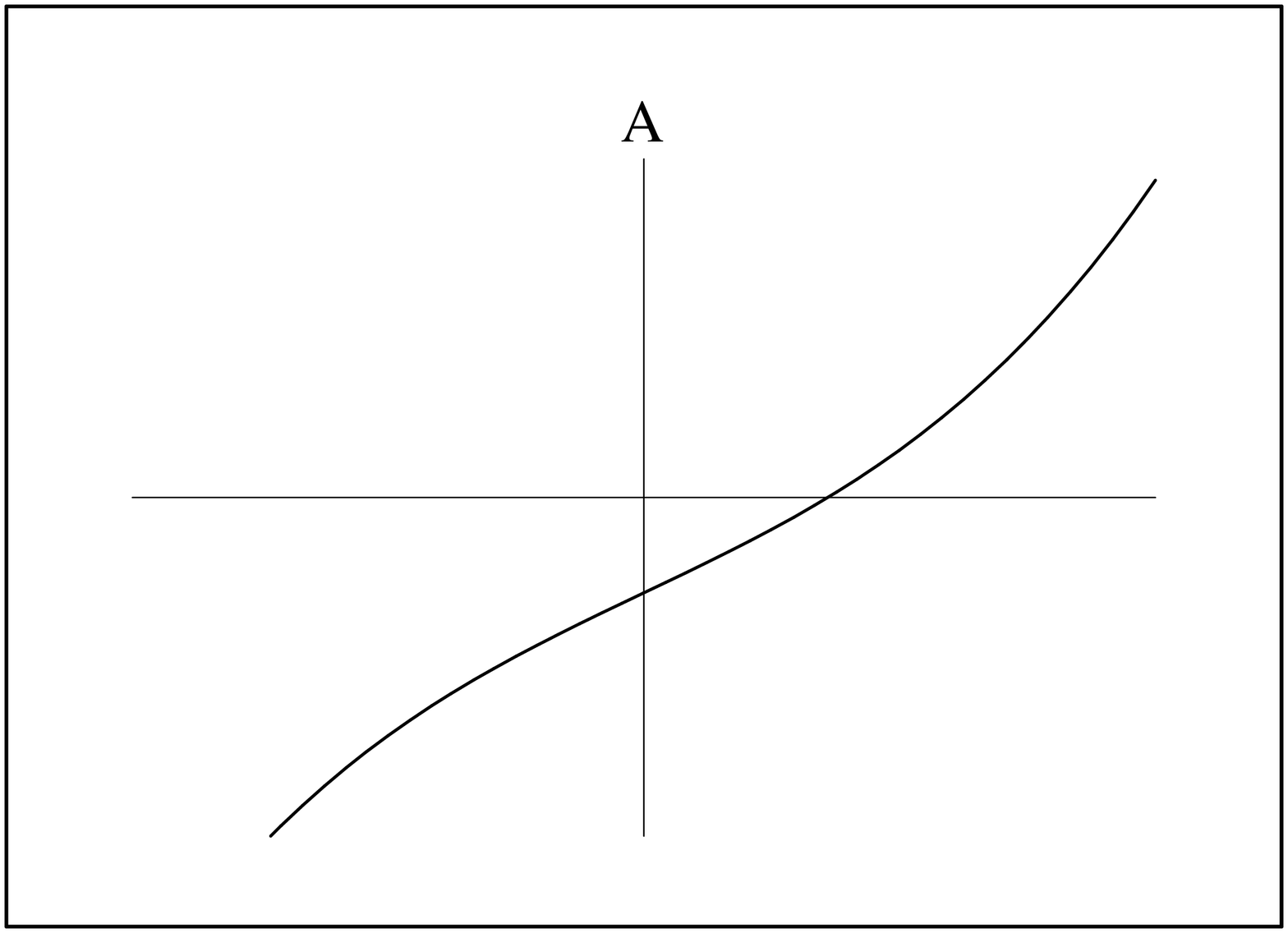, width=4.1cm, height=4.1cm, angle=270}
\epsfig{file=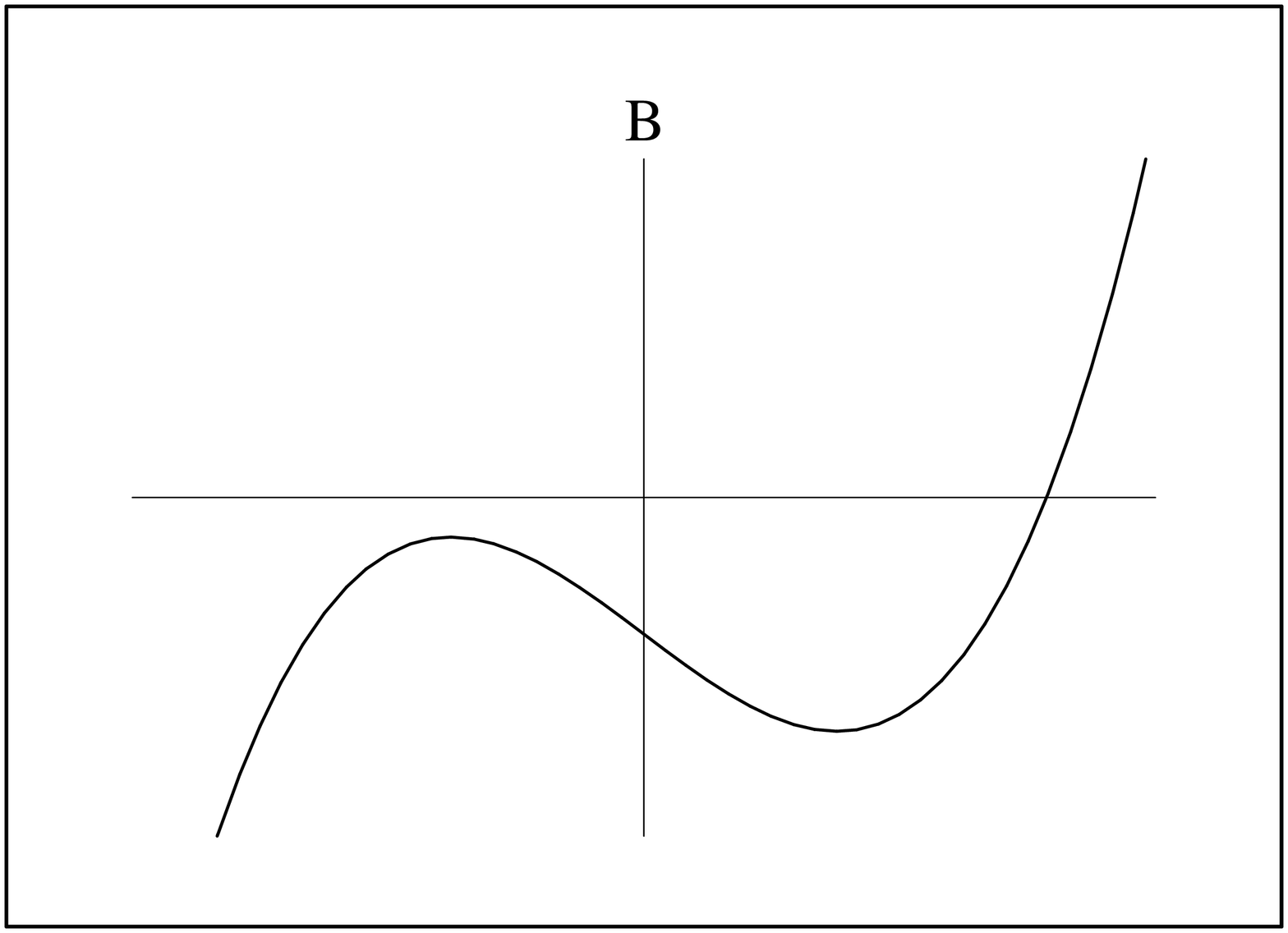, width=4.1cm, height=4.1cm, angle=270} 
\epsfig{file=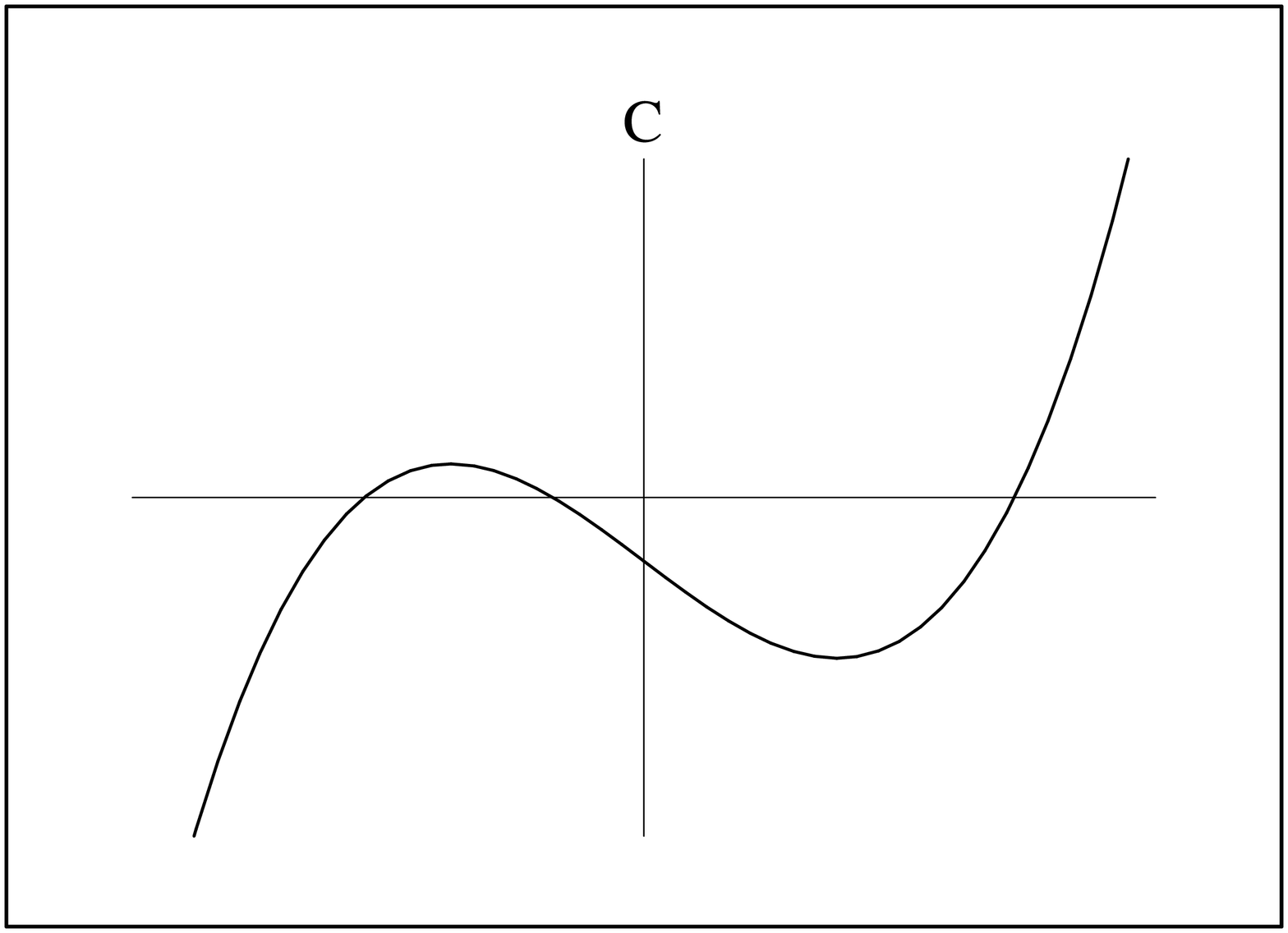, width=4.1cm, height=4.1cm, angle=270}\\
\caption{Graphs of the cubic
$\varrho^3+\varrho(\mu^2-m^2)-\beta^2$ when (A) $\mu>m$, (B) $\mu<m$ and
$\beta^2>\frac{2}{27}(3(m^2-\mu^2))^{3/2}$ and (C) $\mu<m$ and
$\beta^2<\frac{2}{27}(3(m^2-\mu^2))^{3/2}$.}
\end{center}
\end{figure}
The condition on the parameters of the theory required for the bound state
corresponding to this solution to be time oscillatory 
(and for the Hamiltonian to be bounded below) \eqref{0fcond} becomes
\begin{equation}
\label{fcond}
\frac{\beta^2}{\mu^2} \leq m \ .
\end{equation}
We note that when $m=0$ equation \eqref{fcond} can only be satisfied by $\beta=0$,
i.e. a massless field in the bulk can only be coupled to a boundary oscillator in the
presence of a (repulsive) Robin boundary potential. 

In the quantized system the reflection matrix \eqref{0Rmatrix} simplifies to
\begin{equation}
\label{Rmatrix}
R(p)=\frac{(p^2-\mu^2+m^2)p - i \beta^2}
             {(p^2-\mu^2+m^2)p + i \beta^2} \ .
\end{equation}
The reflection matrix \eqref{Rmatrix} has three poles, corresponding to the three
complex roots of \eqref{0cubic}, one of these poles always occurs at
positive, purely imaginary, values of momentum ($p = i \varrho$) and
corresponds to the classical bound state of the system, the features of
which have been discussed in previously.
The remaining two
poles occur at values of momentum in the lower half complex
plane.
To help interpret these poles we will initially assume that $\mu > m$, in which
case the poles are arranged as shown in Figure 11. The dotted
circle describes the locus of points with the same modulus as the
location of the bound state pole.

\begin{figure}
\begin{center}
\begin{pspicture}(-3,-1.5)(3,1.5)
\psline(-3,0)(3,0)
\psline(0,-1.5)(0,1.5)
\rput(2.7,1.3){$p$}
\psline(2.5,1)(2.5,1.5)
\psline(2.5,1)(3,1)
\pscircle[linestyle=dotted](0,0){1}
\pscircle[fillstyle=solid,fillcolor=white](0,-1){0.15}
\pscircle[fillstyle=solid,fillcolor=black](0,1){0.15}
\pscircle[fillstyle=solid,fillcolor=white](-2,0.5){0.15}
\pscircle[fillstyle=solid,fillcolor=black](-2,-0.5){0.15}
\pscircle[fillstyle=solid,fillcolor=white](2,0.5){0.15}
\pscircle[fillstyle=solid,fillcolor=black](2,-0.5){0.15}
\end{pspicture}\\
\caption{The analytic structure of
the reflection matrix of the dynamic boundary for $\mu>m$, $\lambda=0$.}
\end{center}
\end{figure}
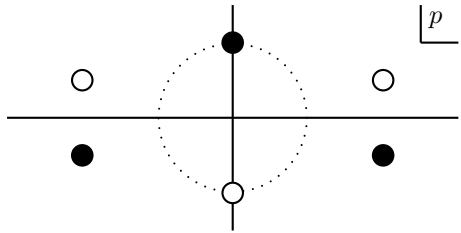

Figure 3A shows the shape of $\sigma_{\text{tot}}$ for the dynamic
boundary in the region of the parameter space where $\mu>m$ and $\lambda=0$.
Figure 3A exhibits a resonance peak,
the pole in the lower right quadrant of
figure 11 corresponds to the resonance state.

We now have a full description of two of the poles of $R(p)$
in the region where $\mu>m$. The third pole of $R(p)$ carries a similar
interpretation to that just presented for the resonance. If we were to
continue figure 3A into
negative values for $\rho$ we would observe a second resonance peak
corresponding to the third pole of the reflection matrix. However, as
mentioned before this region of the reflection matrix, and thus the
cross section is never probed.

Before we conclude our discussion of this region of the parameter space
let us consider the behaviour of the poles of the reflection matrix in
the limit $\beta\to0$, where the oscillator decouples from the field. 
From \eqref{Rmatrix} we see that in this limit
the pole corresponding to the bound state approaches $\varrho=0$ from
above. The energy of this state in the $\beta\to0$ limit is equal to the
mass of the bulk particle, $m$.
This is the same behaviour as for the boundary bound state of the
attractive Robin boundary potential in the limit $\lambda\to0$ and
indicates the bound state arises from the stationary bulk particle
state. The ability for a boundary to support such a state indicates it
is attractive to the particles of the field.
In the limit $\beta\to0$ the poles associated with the resonances
stabilise and have energy equal to the mass of the particles of the
boundary oscillator, $\mu$. This further supports the interpretation of
the resonance particles as particles of the boundary field.

Let us now consider the region of the parameter space where $\mu<m$. 
Figure 12 show the arrangement of poles and nodes of the reflection
matrix in this region. 
\begin{figure}
\begin{center}
\begin{pspicture}(-2,-2)(2,2)
\psline(-2,0)(2,0)
\psline(0,-2)(0,1.75)
\rput(1.75,1.75){$p$}
\psline(1.5,1.5)(1.5,2)
\psline(1.5,1.5)(2,1.5)
\rput(0,2){A}
\pscircle[linestyle=dotted](0,0){1.5}
\pscircle[fillstyle=solid,fillcolor=white](0,-1.5){0.15}
\pscircle[fillstyle=solid,fillcolor=black](0,1.5){0.15}
\pscircle[fillstyle=solid,fillcolor=white](-0.75,0.5){0.15}
\pscircle[fillstyle=solid,fillcolor=black](-0.75,-0.5){0.15}
\pscircle[fillstyle=solid,fillcolor=white](0.75,0.5){0.15}
\pscircle[fillstyle=solid,fillcolor=black](0.75,-0.5){0.15}
\end{pspicture}
\qquad
\begin{pspicture}(-2,-2)(2,2)
\psline(-2,0)(2,0)
\psline(0,-2)(0,1.75)
\rput(1.75,1.75){$p$}
\psline(1.5,1.5)(1.5,2)
\psline(1.5,1.5)(2,1.5)
\rput(0,2){B}
\pscircle[linestyle=dotted](0,0){1.5}
\pscircle[fillstyle=solid,fillcolor=white](0,-1.5){0.15}
\pscircle[fillstyle=solid,fillcolor=black](0,1.5){0.15}
\pscircle[fillstyle=solid,fillcolor=white](0,0.4){0.15}
\pscircle[fillstyle=solid,fillcolor=black](0,-0.4){0.15}
\pscircle[fillstyle=solid,fillcolor=white](0,0.9){0.15}
\pscircle[fillstyle=solid,fillcolor=black](0,-0.9){0.15}
\end{pspicture}\\
\caption{The analytic structure of the reflection matrix for the dynamic boundary for (A)
$\mu<m$ and $\beta^2>\frac{2}{27}(3(m^2-\mu^2))^{3/2}$ and (B) $\mu<m$
and $\beta^2<\frac{2}{27}(3(m^2-\mu^2))^{3/2}$.}
\end{center} 
\end{figure}
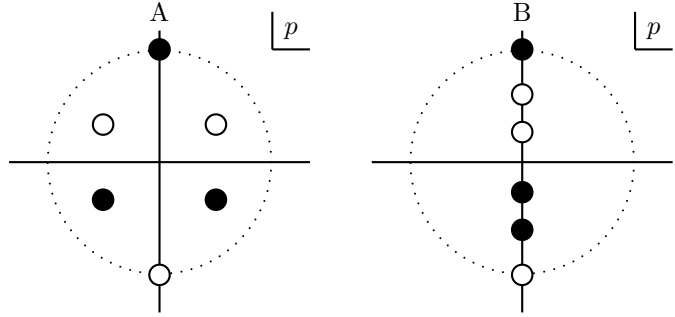

The reflection cross section can be calculated and plotted as
previously, however no additional features beyond background
effects at low momentum are observed. To interpret these poles let us
observe their behaviour as $\beta\to0$. When $\mu<m$ we find that the
bound state pole, being the one located at $p=i\varrho$ on the positive imaginary axis,
no longer goes to $\varrho=0$ as the boundary coupling parameter is
reduced. Instead its limiting energy is $\mu$, the mass of the boundary
particles. As in the previous case there is also a second pole which
also has an energy $\mu$ in the weak coupling limit, but as it is
located on the negative imaginary axis it carries no physical
meaning. The final pole approaches $\varrho=0$ from below as $\beta$ is
reduced, this is the same behaviour as the pole in the reflection matrix
of the repulsive Robin boundary. In both cases the pole does not
correspond to any kind of bound state.

Clearly the spectrum of boundary states of the dynamic boundary is very
different in these two regions of the parameter space. For $\mu>m$ there
is a resonance state corresponding to the particles of the boundary
field and a Robin type boundary bound state arising from the attraction
of the bulk particles to the boundary. However, when $\mu$ is reduced
past $m$ the Robin type bound state vanishes, just as for the Robin
boundary when $\lambda$ is increased past zero. The resonance state
becomes a new, stable, boundary bound state, classically this state can be
thought of as the oscillator driving the field at the boundary but at a
frequency too low to allow the energy to be dissipated away. The absence
of a Robin type boundary bound state suggests that the boundary
oscillator in this region is repulsive to the particles of the field.

\section{Conclusion}

In this paper we considered
the massive Klein-Gordon field coupled to a boundary
oscillator, with and without an additional Robin boundary potential.
We showed that these systems can be solved classically and that the
classical solutions can be written as a superposition of independent
modes of oscillation. 
We observed classical boundary bound state solutions in some regions of the
parameter space, the requirement that these solutions be time oscillatory
is seen to be the same as the requirement that the Hamiltonian is
bounded below. We develop a condition relating the parameters of the theories
which, when satisfied, ensures that bound state is oscillatory and thus that
the system is physical.

Writing the classical solutions as superposition of independent modes
allows the system to be quantized using
canonical methods.
The quantum reflection matrices for
the boundaries are found from the two point function of the field.
We observed several interesting features in the total cross section of the
reflection process, including resonances and Ramsauer-Townsend effects.
Although resonances associated with boundaries have previously been
observed \cite{GZ,MMR,BPT,BBC} they have only previously been associated with boundaries
containing additional degrees of freedom in \cite{BBC}. Ramsauer-Townsend effects
have not previously been observed in boundary field theories.

Quantum boundary bound states were observed in several regions of the
parameter space. Some of these bound states were seen to arise from the
states of the boundary oscillator, this behaviour has not previously
been observed. Other boundary bound states arise in a similar manner to
those of the attractive Robin boundary.

Similar systems to the ones described in this paper have
previously been investigated. The coupling of one or more harmonic
oscillators at fixed points in the bulk of a 1+1 dimensional, massless,
classical Klein-Gordon field is discussed in \cite{Cho}, and
the coupling of an harmonic oscillator to a massless
scalar field in a spherical reflecting cavity
is discussed in \cite{AMMn,FhMM} where it is used to model the behaviour of an excited atom in a cavity. 
In \cite{BBC} the massless scalar field coupled to a boundary oscillator
is studied as a model of brane-bulk interaction in a braneworld universe.
There also exist several papers
dealing with quantum mechanical systems linked the some dissipative medium, based on the
appoach of \cite{CL} where the dissapative medium is modeled by coupling the system to an
infinite number of harmonic oscillators. This modelling of the dispersive medium should be
equivalent to the coupling of the system to the Klein-Gordon field, which contains an infinite
number of harmonic oscillators, however, a correspondence between the results of this paper and those 
from dissapative quantum mechanics has yet to be found. For a recent treatment of the harmonic oscillator
using the methods of \cite{CL} see \cite{Rose}.

It seems that similar results to those presented in this paper should be
obtainable for fields bounded by arbitrary linear mechanical systems. Of particular
interest would be the existence and energy of boundary bound states and
resonances for such systems. Other
possibly interesting extensions include coupling two or more
non-interacting bulk fields through a boundary or reformulating the
present system to describe a point interaction on the whole line, or in
a volume. In the latter case we would expect to make contact with the
results of the earlier papers mentioned above.

\section*{Acknowledgments}

The author would like to thank Gustav Delius, Zolt\`{a}n Bajnok, L\'{a}szl\'{o} Palla,
G\'{a}bor Tak\'{a}cs, Evgeni
Sklyanin and Bernard Kay for useful and interesting discussions. The
author was supported by a PPARC studentship whilst undertaking this research.

\end{document}